# The mechanics of solid-state nanofoaming

Frederik Van Loock[1], Victoria Bernardo[2], Miguel Angel Rodríguez Pérez[2], Norman A. Fleck[1*]

1. Engineering Department, University of Cambridge, Trumpington Street, CB2 1PZ Cambridge, United Kingdom

2. Cellular Materials Laboratory (CellMat), Condensed Matter Physics Department, University of Valladolid, Paseo de Belen 7, 47011, Valladolid, Spain

*Corresponding Author

E-mail address: naf1@eng.cam.ac.uk (N.A. Fleck)

**Abstract**

Solid-state nanofoaming experiments are conducted on two PMMA grades of markedly different molecular weight using $CO_2$ as the blowing agent. The dependence of porosity of the nanofoams upon foaming time and foaming temperature is measured. Also, the microstructure of the PMMA nanofoams is characterized in terms of cell size and cell nucleation density. A one dimensional numerical model is developed to predict the growth of spherical, gas-filled voids during the solid-state foaming process. Diffusion of $CO_2$ within the PMMA matrix is sufficiently rapid for the concentration of $CO_2$ to remain almost uniform spatially. The foaming model makes use of experimentally calibrated constitutive laws for the PMMA grades, and the effect of dissolved $CO_2$ is accounted for by a shift in the glass transition temperature of the PMMA. The observed limit of achievable porosity is interpreted in terms of cell wall tearing; it is deduced that the failure criterion is sensitive to cell wall thickness.

## 1. Introduction

Polymeric nanofoams are polymer foams with an average cell size of below 1 micrometer [1]. This relatively new class of porous solids has the potential to offer unique and attractive combinations of thermal, mechanical, and optical properties [2–4]. For example, the thermal conductivity of polymeric nanofoams can be lower than that of air (= 0.025 W m$^{-1}$K$^{-1}$). When the average cell size is in the order of the mean free path of the gas molecules in the cells (close to 70 nm for air at standard conditions), the thermal conductivity of the gas in the foam is significantly reduced due to the Knudsen effect [5,6].

Wang *et al.* [7] calculated that a polymeric nanofoam has a thermal conductivity close to or below 0.025 W m$^{-1}$K$^{-1}$ when the average cell size $l$ is below 200 nm and the porosity $f$ exceeds 0.85. To achieve this morphology, the cell nucleation density $N_d$ must exceed $10^{21}$ m$^{-3}$ [1]. A large number of experimental studies focus on the effect of processing conditions and the choice of polymer precursor upon the cell nucleation density $N_d$, the void size $l$ and the porosity $f$ of polymeric nanofoams, as reviewed by Costeux [1]. Many of these studies make use of the solid-state foaming method in which a physical blowing agent (e.g. $CO_2$) is used to nucleate and grow cells in a polymer matrix such as polymethyl methacrylate (PMMA) [8,9]. The review by Costeux [1] on experimental studies of solid-state nanofoaming discusses the trade-off between porosity and cell size of a polymeric nanofoam. Polymeric nanofoams of $l <$ 200 nm are reported for a nucleation density above $10^{21}$ m$^{-3}$, but their porosity is limited to close to 0.85, see, for example, Aher *et al.* [10] and Costeux and Zhu [11]. The observed porosity limit for nanofoams with a nucleation density above $10^{21}$ m$^{-3}$ may be due to the fact that the walls between the nano-sized cells are limited by the end-to-end distance of the individual polymer chains [1,12]. Polymeric nanofoams of porosity on the order of 0.8 to 0.9 have been produced, but their cell size is well above 200 nm (and $N_d << 10^{21}$ m$^{-3}$) [1,9,13]. The microstructural requirement for polymeric nanofoams with a thermal conductivity lower than the thermal conductivity of air is currently beyond the practical limit of the state-of-the-art solid-state nanofoaming process [7].

The final porosity and final cell size in solid-state nanofoaming can be predicted by simulating void growth. In contrast to the substantial body of experimental work on polymeric nanofoams produced by solid-state foaming, as reviewed by Costeux [1], and the development of cell growth models for liquid state foaming processes [14–16], theoretical studies on cell growth during solid-state nanofoaming are limited. Costeux and co-workers [13,17] simulated void nucleation and void growth during the solid-state nanofoaming of acrylate co-polymers by making use of the model of Shafi *et al.* [18]. They conducted a series of nanofoaming experiments but their model overestimates the measured final porosity of the nanofoams. The mismatch between the simulated and the measured porosity of acrylic nanofoams may be due to (i) the assumption that cell growth continues until the foaming temperature attains the glass transition temperature of the polymer-gas solid and/or (ii) the assumption that the polymer-gas solid surrounding the cell is in a liquid (viscous) state throughout the solid-state foaming process. In reality, void growth occurs at a temperature above and below the glass transition temperature of the solid surrounding the void. This is addressed in detail in the present study.

*Scope of study*

PMMA nanofoams are produced from two PMMA grades of widely different molecular weight; a solid-state foaming process is used with $CO_2$ as the blowing agent. We characterize the microstructure of the nanofoams in terms of porosity $f$, cell size $l$, and cell nucleation density $N_d$. In addition, we develop a void growth model, based on the constitutive law of PMMA grades close to the glass transition temperature, by building on the recent study of Van Loock and Fleck [19]. Both predicted and measured final porosities are obtained as a function of foaming time and foaming temperature; also, cell wall tearing mechanisms are considered in order to account for the observed limit in final porosity.

## 2. Nanofoaming experiments

### 2.1 Materials

Foaming experiments were conducted on two PMMA grades: pelletized PMMA (Altuglas V825T) of average molecular weight[1] $M_w$ = 92 500 g mol$^{-1}$ and cast PMMA sheets (Altuglas CN with sheet thickness close to 3 mm) with $M_w$ = 3 580 000 g mol$^{-1}$. We shall refer to the Altuglas V825T and Altuglas CN grades as 'low $M_w$ PMMA' and 'high $M_w$ PMMA', respectively. Both grades have a density $\rho^p$ equal to 1 190 kg m$^{-3}$ (as measured at 23 °C and at 50% relative humidity). The glass transition temperature ($T_g$ = 114.5 °C) of the low $M_w$

---

[1] The average molecular weight was measured by gas permeation chromatography (GPC) with an Agilent Technologies PL GPC220 (USA) instrument with a nominal flow rate equal to $1.67 \times 10^{-5}$ l s$^{-1}$ at a test temperature equal to 30 °C.

PMMA is close to the glass transition temperature ($T_g$ = 116.5 °C) of the high $M_w$ PMMA as measured by differential scanning calorimetry (DSC) using a heating rate of 10 °C min$^{-1}$.

**2.2 Solid-state nanofoaming experiments**

Foaming precursors of the low $M_w$ and high $M_w$ PMMA grades were made as follows. The low $M_w$ PMMA pellets were heated to 250 °C for 450 s, then compressed for 60 s between two heated plates at a pressure equal to 17 MPa. The resulting sheet was cooled to room temperature with the pressure of 17 MPa maintained. Cuboid precursors of dimension 20x10x3 mm$^3$ were machined from the low $M_w$ PMMA sheet and from the as-received high $M_w$ PMMA sheet.

Foaming experiments were performed using a pressure vessel[2] with a pressure controller[3] and temperature controller[4]. Medical grade $CO_2$ (> 99.9% purity) was used as the blowing agent for the foaming experiments. The two step solid-state foaming process was employed, as detailed in the study of Martin-de León *et al.* [9]. The precursor samples were held in the pressure vessel at a constant $CO_2$ saturation pressure equal to 31 MPa, and at a constant temperature equal to 25 °C for 24 hours in order to ensure saturation of the $CO_2$ into the PMMA. The mass concentration[5] C, at equilibrium, is close to 24 wt% for both the low and high $M_w$ PMMA, according to the measurement procedure detailed by Martin-de León *et al.* [9]. Next, the pressure was progressively released to atmospheric pressure with an instantaneous pressure drop rate close to 100 MPa s$^{-1}$. The samples were then foamed in a foaming bath[6] at selected foaming temperatures (25 °C, 40 °C, 60 °C, 80 °C, 100 °C) and selected foaming times[7] (60 s, 180 s, 300 s, and 600 s). It is assumed in the remainder of the study that the foaming times are sufficiently long for the temperature to be spatially uniform[8] within the sample.

---

[2] Pressure vessel model PARR 4681 of Parr Instrument Company (USA).
[3] Pressure controller pump SFT-10 of Supercritical Fluid Technologies Inc (USA).
[4] Temperature controller CAL 3300 of CAL Controls Ltd (UK).
[5] We define the mass concentration $C$ of $CO_2$ in PMMA with respect to the total mass of the PMMA-$CO_2$ mixture. Note that the definition of $CO_2$ solubility (with respect to the mass of the PMMA absent $CO_2$) is used in the work of Martin-de León *et al.* [9].
[6] Thermal bath J.P. Selecta Model 6000685 of Grupo Selecta (Spain). The time between the pressure release and the start of foaming was close to 120 s.
[7] Samples were immersed in a water bath at a temperature close to 10 °C at the end of the foaming time.
[8] The justification for this assumption is as follows. Immersion of the sample in water or oil provides excellent heat transfer at the surface of the sample. The time constant $\tau = x^2 / \kappa \approx$

## 2.3 Characterization of the PMMA nanofoams

*Porosity*

The density $\rho^f$ of the foamed samples was determined by the water-displacement method with a weight balance[9]. A surface layer of depth 200 μm was removed by polishing[10] to ensure that the solid skin (of thickness below 100 μm) was absent before the density measurements were made. The porosity $f$ of the samples is obtained by:

$$f = 1 - \frac{\rho^f}{\rho^p} \qquad (1)$$

where $\rho^p$ ($= 1\,190$ kg m$^{-3}$) is the density of solid PMMA.

*Microstructure*

Foamed samples were cooled in liquid nitrogen and then fractured. The fracture surfaces were coated with a layer of gold by sputtering[11], and micrographs of the coated fracture surfaces were taken by a scanning electron microscope[12] (SEM). The cellular structure of each material was characterised by analysing the micrographs with dedicated in-house software based on ImageJ/FIJI [20]. Microstructural parameters such as the average cell size $l$, standard deviation $s$ of the observed cell sizes, and cell nucleation density $N_d$, using the method as suggested by Kumar and Suh [21], were obtained[13].

*Open cell content*

The open cell content of the foamed samples was measured by gas pycnometry[14] with nitrogen in accordance with the ASTM D6226-15 standard [23]. The open cell content ratio $O_v$ is

---

20 s where $x = 1.5$ mm is the half-thickness of the PMMA sample and $\kappa = 1.1 \times 10^{-7}$ m$^2$ s$^{-1}$ is the thermal diffusivity of PMMA at room temperature [22].
[9] Analytical balance AT261 of Mettler-Toledo (USA).
[10] Grinding and polishing system LaboPOl2-LaboForce3 of Struers (USA).
[11] Sputter coater SDC 005 of Balzers Union (Liechtenstein).
[12] Scanning electron microscope QUANTA 200 FEG of Thermo Fisher Scientific (USA).
[13] At least 200 cells were analysed from multiple micrographs per foamed sample.
[14] Gas pycnometer (USA) AccuPyc II 1340 of Micromeritics (USA).

defined as the ratio of the volume of open pores to the total pore volume of a foam, and is obtained by:

$$O_v = \frac{V^g - V^p - V^s}{fV^g} \quad (2)$$

where $V^g$ is the geometric volume of the foam, $V^p$ is the pycnometer volume and $V^s$ is a penalty volume to account for the volume of the cells at the surface of the foam. The penalty volume $V^s$ is assumed to be close to zero in the case of nanofoams. The geometric volume $V^g$ is measured by the water-displacement method as detailed above. Foamed samples were subjected to a pressure scan from 0.02 MPa to 0.13 MPa in the gas pycnometer. The pycnometer volume initially decreases as the gas pressure increases until the interconnected open cells are completely filled with gas and the pycnometer volume remains constant at increased pressures. We take this constant value of pycnometer volume $V^p$ in order to calculate $O_v$ via Eq. (2).

## 3. Results nanofoaming experiments

The measured porosity $f$, average observed cell size $l$, standard deviation $s$ of observed cell sizes, and cell nucleation density $N_d$ of the nanofoams are reported in Tables 1 and 2 for the low $M_w$ and high $M_w$ grades of PMMA, respectively. In addition, a representative series of SEM micrographs of the nanofoams are shown in Figure 1. The low $M_w$ and the high $M_w$ nanofoams have contrasting microstructures and the cell nucleation density of the low $M_w$ nanofoams ($N_d \approx 2 \times 10^{20} \, \text{m}^{-3}$) is an order of magnitude less than that of the high $M_w$ nanofoams ($N_d \approx 2 \times 10^{21} \, \text{m}^{-3}$). The average cell size $l$ of the high $M_w$ nanofoams ranges from 20 nm to 50 nm, and is an order of magnitude smaller than the average cell size of the low $M_w$ nanofoams (of size 200 nm to 350 nm). These values of $l$ and $N_d$ for the low $M_w$ nanofoams are consistent with the results of Martin-de León *et al.* [9] who conducted solid-state foaming experiments with an identical low $M_w$ PMMA grade. The measured average cell size $l$ of the low $M_w$ and the high $M_w$ nanofoams, as a function of foaming time $t_f$ for $T_f = 60\,°C$, is plotted in Figure 2a. Void growth typically occurs over a foaming time period of 60 s to 180 s, followed by arrest. There is a mild dependence of the foaming temperature $T_f$ upon the final value for $l$, see Tables 1 and 2.

The measured porosity $f$ of the nanofoams is plotted as a function of $t_f$ in Figure 2b for $T_f = 60\,°C$ and for $T_f = 100\,°C$. Consistent with the $l$ versus $t_f$ curves for $T_f = 60\,°C$, as presented in Figure 2a, the porosity increases over a foaming period of 60 s to 180 s until a stable ($t_f$-independent) value of final porosity is achieved. The highest observed porosity of the low $M_w$ PMMA nanofoams ($f_{max} = 0.75$) is approximately 25% higher than that of the high $M_w$ PMMA nanofoams ($f_{max} = 0.60$). At a foaming temperature of $T_f = 100\,°C$, the porosity decreases with increasing foaming time beyond $t_f = 60$ s, and this is due to collapse of the foamed structure. This behavior is also illustrated in plots of $f$ versus $T_f$, over the explored range of foaming times, see Figures 2c and 2d for the low $M_w$ and high $M_w$ PMMA nanofoams, respectively.

The measured open cell content $O_v$ is plotted as a function of the measured porosity $f$ in Figure 3a (low $M_w$) and in Figure 3b (high $M_w$) for $20\,°C \leq T_f \leq 80\,°C$. Nanofoams with porosities well below the highest observed porosity $f_{max}$ are closed-cell in nature. An abrupt transition to an open-celled structure occurs close to $f_{max}$. The observed collapse of the foam at $T_f = 100\,°C$ is preceded by cell wall failure for the low $M_w$ nanofoams (see Figure 1b) and by the formation of cracks interconnecting the nano-sized pores in for the high $M_w$ nanofoams (see Figure 1d).

**4. Void growth model**

A void growth model is now developed to predict porosity as a function of foaming time and foaming temperature for the PMMA nanofoams. The expansion of a pre-existing as-nucleated cavity during solid-state nanofoaming is simulated by means of a one dimensional single cell growth model [15,24]. A finite shell surrounds the void in order to account for void-void interaction in an approximate manner. More sophisticated models of an array of voids (such as periodic cell models) could be adopted but the intent here is to emphasize the strong role of the evolving constitutive response. Consider a polymer-gas solid with equisized spherical voids. A cross-section of the undeformed (reference) configuration of the spherical void, with initial radius $a_0$ and initial outer radius $b_0$, along with the adopted spherical coordinate system $(r,\theta,\phi)$, is shown in Figure 4. Assume that the initial gas pressure $p_0$ in the as-nucleated void equals the saturation pressure during the saturation phase prior to nucleation of the voids. The deformed configuration for the void of inner radius $a$ and outer radius $b$ at time $t$ is shown in Figure 4.

*Kinematics*

Assume that the void remains spherical during growth and that the solid surrounding the void is incompressible. Then a material point, initially at radius $R$, is displaced to a radius $r$ such that:

$$r^3 - a^3 = R^3 - a_0^3 \tag{3}$$

by incompressibility. For later use, this relation is re-arranged to the form:

$$\left(\frac{r}{R}\right)^3 = 1 + \left(\frac{a_0}{R}\right)^3\left[\left(\frac{a}{a_0}\right)^3 - 1\right] \tag{4}$$

Note that $r/R$ is a function of the time-like variable ($a/a_0$) and of the Lagrangian position variable $R/a_0$. The von Mises effective strain $\varepsilon_e$ is defined in the usual manner as $\varepsilon_e^2 = \frac{2}{3}\varepsilon_{ij}\varepsilon_{ij}$, giving:

$$\varepsilon_e = |2\varepsilon_{\theta\theta}| = 2\ln\left(\frac{r}{R}\right) \tag{5}$$

where $\varepsilon_{\theta\theta}$ is the hoop strain. Now insert Eq. (4) into Eq. (5) to obtain:

$$\varepsilon_e = \frac{2}{3}\ln\left[1 + \left(\frac{a_0}{R}\right)^3\left(\left(\frac{a}{a_0}\right)^3 - 1\right)\right] \tag{6}$$

and take the time derivative of $r$ in Eq. (3) to give:

$$\dot{r} = v_r = \left(\frac{a}{r}\right)^2 \dot{a} \tag{7}$$

where $v_r$ is the radial velocity of a material element at $r$. Consequently, the effective strain rate $\dot{\varepsilon}_e$ reads:

$$\dot{\varepsilon}_e = \left|\frac{\partial v_r}{\partial r}\right| = \frac{2a^2}{R^3}\left(\frac{r}{R}\right)^{-3}\dot{a} \tag{8}$$

*Equilibrium*

Write ($\sigma_{rr}$, $\sigma_{\theta\theta}$, $\sigma_{\phi\phi}$) as the active stress components in the spherical coordinate system. Radial equilibrium dictates that [25]:

$$\frac{\partial \sigma_{rr}}{\partial r} + \frac{1}{r}\left(2\sigma_{rr} - \sigma_{\theta\theta} - \sigma_{\phi\phi}\right) = 0 \tag{9}$$

Due to symmetry, $\sigma_{\phi\phi} = \sigma_{\theta\theta}$ and Eq. (9) simplifies to:

$$\frac{\partial \sigma_{rr}}{\partial r} = \frac{2(\sigma_{\theta\theta} - \sigma_{rr})}{r} = \frac{2\sigma_e}{r} \tag{10}$$

where $\sigma_e = \sigma_{\theta\theta} - \sigma_{rr}$ is the von Mises effective stress [26]. Integration of Eq. (10) leads to:

$$p - p_a = \int_{r=b}^{r=a} \frac{2\sigma_e}{r}\,\mathrm{d}r \tag{11}$$

where $p$ is the gas pressure inside the cavity for a given radius $a$, and $p_a$ is the ambient pressure. Upon making use of Eq. (3), the above integral can be re-phrased as:

$$p - p_a = \int_{R=b_0}^{R=a_0} \frac{2}{R}\left(\frac{R}{r}\right)^3 \sigma_e\,\mathrm{d}R \tag{12}$$

The effective stress $\sigma_e$ is a function of the effective strain $\varepsilon_e$, the effective strain rate $\dot{\varepsilon}_e$, and the normalized temperature $T/T_g$ via the constitutive law for the PMMA-$CO_2$ solid, of general functional form F where:

$$\sigma_e = \mathrm{F}\left(\varepsilon_e, \dot{\varepsilon}_e, T/T_g\right) \tag{13}$$

The choice of F is given below. We emphasize that the glass transition temperature $T_g$ of the PMMA is a function of $CO_2$ concentration and that the effective strate rate $\dot{\varepsilon}_e$ scales with $\dot{a}$ as prescribed in Eq. (8). Upon substituting Eq. (13) into Eq. (12) we obtain an expression for $\dot{a}$, and we integrate over time in order to determine the time evolution of $a/a_0$.

*Mass conservation*

At the start of the foaming process, the chemical potential of the $CO_2$ molecules in the nucleated voids is lower than chemical potential of $CO_2$ molecules in the PMMA-$CO_2$ solid. Consequently, $CO_2$ gas molecules migrate from the PMMA-$CO_2$ solid into the voids. The concentration of $CO_2$ gas molecules $C(r,t)$ at time $t$ and position r (for $a < r < b$) can be obtained by solving Fick's second law of diffusion [27]:

$$\frac{\partial C}{\partial t} = \frac{D}{r^2}\frac{\partial}{\partial r}\left[r^2 \frac{\partial C}{\partial r}\right] \tag{14}$$

in the deformed configuration, where $D$ is the diffusion coefficient for $CO_2$ in PMMA. Measurements of $D$ at temperatures and pressures typical for solid-state nanofoaming of PMMA by $CO_2$ are available in the literature as follows. Guo and Kumar [28] measured $D$ based on desorption measurements and found that $D$ ranges from $D = 2.5 \times 10^{-12}$ m² s⁻¹ to $D = 3.65 \times 10^{-11}$ m² s⁻¹ for temperatures ranging from -30°C to 100 °C at a $CO_2$ pressure equal to 5 MPa. Li *et al.* [29] measured $D$ by a sorption technique and found that $D$ is in the range of $6 \times 10^{-11}$ m² s⁻¹ to $9.5 \times 10^{-11}$ m² s⁻¹ for temperatures ranging from 30°C to 70 °C, and pressures ranging from 6 MPa to 18 MPa. Now, introduce a characteristic diffusion time $\tau_D$:

$$\tau_D = \frac{(L_D)^2}{D} \tag{15}$$

where $L_D$ is a diffusion length which is approximated for the void growth problem by:

$$L_D \approx b_0 \approx \left(\frac{3}{4\pi N_d}\right)^{\frac{1}{3}} \tag{16}$$

Observations of cell nucleation densities of PMMA nanofoams ($N_d > 10^{20}$ m⁻³) imply $L_D <$ 133 nm [1]. Upon assuming $D = 10^{-12}$ m²/s, we obtain $\tau_D \approx$ 20 ms via Eq. (15), which is two orders of magnitude lower than typical observed cell growth times for solid-state nanofoaming of PMMA by $CO_2$ as reported by Martín-de León *et al.* [9] and reported above. We conclude that the $CO_2$ concentration profile $C(R,t)$ is spatially uniform at all times: $C(R,t) = C(t)$. Consequently, we do not need to solve the diffusion equation to predict void growth during solid-state nanofoaming of PMMA by $CO_2$.

We shall assume that the mass of gas molecules in the voids and in the surrounding solid is constant; leakage of gas molecules to neighbouring voids or the sample's environment is neglected. The resulting mass conservation statement for $CO_2$ reads:

$$C\rho^p \left(b^3 - a^3\right) + \rho^g a^3 = C_0 \rho^p \left(b_0^3 - a_0^3\right) + \rho_0^g a_0^3 \qquad (17)$$

where $\rho^p$ is the density[15] of the PMMA-$CO_2$ solid and $\rho^g$ is the density of the $CO_2$ in the voids.

The relations Eq. (12) and Eq. (17) form a coupled pair of equations which can be solved to obtain $p$ as a function of cavity expansion $a/a_0$. Additional assumptions are detailed below.

*Depression of the glass transition temperature by $CO_2$*

The dissolution of $CO_2$ into a linear, amorphous polymer such as PMMA reduces the glass transition temperature $T_g$ of the PMMA-$CO_2$ solid. This plasticization effect is attributed to the increased mobility of PMMA chains due to lubrication by the $CO_2$ molecules, and the decrease of the intermolecular bond strength, as the $CO_2$ molecules increase the spacing between the PMMA chains [30,31]. A range of experimental techniques have been used in the literature to determine the glass transition temperature $T_g$ of PMMA as a function of $CO_2$ mass concentration $C$. Chiou *et al.* [32] made use of DSC to measure $T_g/T_g^0$ as a function of $C$, where $T_g^0 = T_g(C=0)$. Likewise, Wissinger and Paulaitis [33] measured the dependence of $T_g/T_g^0$ upon $C$ via creep compliance measurements. Guo and Kumar [28] made use of solid-state foaming experiments to observe the relation between $T_g/T_g^0$ and $CO_2$ for a PMMA-$CO_2$ mixture. The measured $T_g/T_g^0$ versus $C$ data, for PMMA-$CO_2$, as reported by Chiou *et al.* [32], Wissinger and Paulaitis [33], and Guo and Kumar [28] are shown in Figure 5. Chow [34] used statistical thermodynamics to predict $T_g/T_g^0$ as a function of $C$ and introduced a parameter $\theta$ where:

$$\theta = \frac{M_w^p}{zM_w^g}\frac{C}{1-C} \tag{18}$$

---

[15] We assume that the density of the PMMA-$CO_2$ solid is equal to the density of PMMA absent $CO_2$ at standard conditions (i.e. $\rho^p = 1190$ kg m$^{-3}$) based on the measurements of Pantoula and Panayiotou [40] and Pantoula *et al.* [41] who observed that the relative increase in volume of a PMMA-$CO_2$ mixture is close to the relative increase of the mass of a PMMA-$CO_2$ mixture for a $CO_2$ pressure up to 30 MPa.

Here, $M_w^p$ is the molecular weight of the polymer repeat unit ($M_w^p$ = 100.12 g mol$^{-1}$ for a methyl methacrylate monomer), $M_w^g$ is the molecular weight of the gas ($M_w^g$ = 44.01 g mol$^{-1}$ for $CO_2$), and $z$ is a lattice coordination number equal to 2, as suggested by Chow [34]. In addition, Chow [34] defined a parameter $\beta$:

$$\beta = \frac{zR}{M_w^p \Delta C_p} \tag{19}$$

where $R$ is the universal gas constant and $\Delta C_p$ is the change in specific heat capacity of the polymer at the glass transition temperature at constant pressure. The normalized glass transition temperature is then predicted by:

$$\frac{T_g}{T_g^0} = \exp\left[\beta\left((1-\theta)\ln(1-\theta) + \theta\ln\theta\right)\right] \tag{20}$$

Equation (20) is curve fitted to the measured $T_g/T_g^0$ versus $C$ data shown in Figure 5 by a suitable choice of $\Delta C_p$. The fitted value for $\Delta C_p$ equals 355 J kg$^{-1}$ K$^{-1}$ which is slightly higher than the value of $\Delta C_p$ for PMMA as measured by DSC, see Chiou *et al.* [32] and Li *et al.* [35].

*Constitutive model for the PMMA-$CO_2$ solid*

We assume that the effective stress $\sigma_e$ of the PMMA-$CO_2$ solid at a given strain $\varepsilon_e$, strain rate $\dot{\varepsilon}_e$ and normalized temperature $T/T_g$ is the same as that given by PMMA in the absence of $CO_2$: the effect of $CO_2$ is accounted for by a shift in the value for $T_g$. The deformation mechanisms for PMMA in uniaxial tension close to the glass transition temperature have been reviewed recently by Van Loock and Fleck [19] an deformation mechanism maps were constructed by performing a series of uniaxial tension tests on the high $M_w$ PMMA over a range of temperatures near the glass transition and over two decades of strain rate. The operative deformation mechanism depends upon the temperature $T/T_g$, the strain rate $\dot{\varepsilon}_e$, and strain $\dot{\varepsilon}_e$. We shall make use of the constitutive models as calibrated by Van Loock and Fleck [19] for the high $M_w$ PMMA: the Ree-Eyring equation and a rubbery-flow model. For the low $M_w$ PMMA it is necessary to construct an alternative deformation mechanism map. This is reported in the Appendix. For this grade, the relevant deformation mechanisms are Ree-Eyring and viscous flow.

The Ree-Eyring equation relates $\sigma_e$ in the glassy and glass transition regime to temperature $T/T_g$ and strain rate $\dot{\varepsilon}_e$:

$$\frac{\dot{\varepsilon}_e}{\dot{\varepsilon}_0} = \sinh\left(\frac{\sigma_e v}{kT}\right)\exp\left(\frac{-q}{kT}\right) \qquad (21)$$

where $\dot{\varepsilon}_0$ is a reference strain rate, $q$ is an activation energy, $v$ is an activation volume, and $k$ is Boltzmann's constant. Visco-elastic effects are neglected in this finite strain regime. Van Loock and Fleck [19] also fitted an empirical equation to relate $\sigma_e$ to $T/T_g$ and $\dot{\varepsilon}_e$ in the rubbery regime for the high $M_w$ PMMA:

$$\sigma_e = E_0\left(1 - \alpha_R \frac{T}{T_g}\right)\left(\frac{\dot{\varepsilon}_e}{\dot{\varepsilon}_R}\right)^n \varepsilon_e \qquad (22)$$

where $E_R^0$ is a reference modulus, $\alpha_R$ is a temperature sensitivity coefficient, $\dot{\varepsilon}_R$ a reference strain rate, and $n$ a strain rate sensitivity coefficient.

Note that the rubbery regime above the glass transition is absent for PMMA grades of relatively low molecular weight, i.e. $M_w < 150$ kg mol$^{-1}$ [36]. Instead, a linear, viscous flow rule can be used to describe the constitutive behavior of a low $M_w$ PMMA for $T/T_g \gg 1$:

$$\sigma_e = 3\eta\dot{\varepsilon}_e \qquad (23)$$

where $\eta$ is a temperature-dependent viscosity [37,38]:

$$\eta = \eta_0 \exp\left(\frac{-C_1(T/T_g - 1)}{C_2/T_g + T/T_g - 1}\right) \quad (24)$$

in terms of a reference viscosity $\eta_0$ at $T/T_g = 1$; $C_1$ and $C_2$ are fitting constants.

The dependence of the effective stress $\sigma_e$ upon normalized temperature $T/T_g$ and strain rate $\dot{\varepsilon}_e$ is assumed to be governed by Eq. (21) and Eq. (22) for the high $M_w$ PMMA and by Eq. (21) and Eq. (23) for the low $M_w$ PMMA. The fitted parameters for the constitutive laws for the high $M_w$ PMMA are taken from Van Loock and Fleck[16] [19] and are summarized in Table 3. An additional series of tensile tests have been performed on the low $M_w$ PMMA at temperatures close to the glass transition in order to calibrate Eq. (21) and Eq. (23) for the low $M_w$ PMMA as detailed in the Appendix. The resulting calibrated parameters for Eq. (21) and Eq. (23) for the low $M_w$ PMMA are included in Table 3.

*Gas laws*

The equilibrium concentration $C$ of $CO_2$ in PMMA is a function of $CO_2$ pressure $p$ and of temperature. Here, we assume that Henry's law suffices such that [39–42]:

$$C = K_H p \quad (25)$$

where Henry's law coefficient $K_H$ is assumed to be independent of both temperature and pressure. Assume that the concentration of $CO_2$ at the surface of the cavity ($R = a_0$) is in equilibrium with the $CO_2$ pressure within the void via Eq. (25). Take $K_H = 7.74 \times 10^{-9}$ Pa$^{-1}$ for both the low $M_w$ and the $M_w$ PMMA grades, based on the measured $C = 0.24$ equilibrium concentration of $CO_2$ in PMMA at a pressure $p$ equal to 31 MPa and temperature $T = 25$ ˚C, as detailed in section 2.2. Also, assume that the $CO_2$ gas in the void statisfies the ideal gas law:

$$p = \frac{\rho^g RT}{M_w^g} \quad (26)$$

*Temperature-time profile during void growth*

---

[16] We assume that the dependence of the effective stress $\sigma_e$ of the PMMA-$CO_2$ solid upon pressure is small as a first order approximation for the void growth problem.

During the rapid release of pressure at the end of the saturation phase, the samples cool down from the saturation temperature equal to 25 ˚C to a temperature[17] $T_0 = -15$ ˚C due to adiabatic cooling of the expanding gas. The samples are subsequently placed in a thermal bath at a maintained foaming temperature $T_f$. Upon submersion in the foaming bath, assume that the temperature profile $T(t)$ is of the form:

$$T = T_0 + (T_f - T_0)(1 - \exp(-t/\tau)) \qquad (27)$$

where $\tau$ is a time constant associated with the heat conduction into the PMMA, as measured by a thermocouple.

*Void growth simulations*

Void growth during solid-state foaming is simulated by solving the equilibrium Eq. (12) and the mass conservation statement Eq. (17) simultaneously, with due account of the dependence of $T_g$ upon $C$ via Eq. (20), the dependence of the effective stress $\sigma_e$ of the PMMA-$CO_2$ solid upon $\varepsilon_e$, $\dot{\varepsilon}_e$ and $T/T_g$ (Eqs. (21) to (23)), the gas laws (Eqs. (26) and (25)), and the time-temperature profile as captured by Eq. (27). The resulting system of equations is solved by numerical integration[18]. The values of the processing parameters and the material properties are summarized in Table 4. Note that the initial porosity $f_0$ is:

$$f_0 = \left(\frac{a_0}{b_0}\right)^3 \qquad (28)$$

and is estimated[19] to equal $10^{-3}$ for both the low $M_w$ and high $M_w$ PMMA nanofoams. The initial void radius $a_0$ is estimated by:

---

[17] Measured by placing a thermocouple on the sample after pressure release at the end of the saturation phase.

[18] The numerical integration was conducted within the Matlab computing environment by means of the *ode15s* function.

[19] The initial porosity $f_0$ is estimated by saturating low $M_w$ and high $M_w$ PMMA precursors with $CO_2$ at $p = 31$ MPa and $T = 25$ °C. Upon release of the pressure to atmospheric pressure, the samples were immediately immersed in liquid nitrogen to prevent the growth of the nucleated voids. The porosity of the samples was measured by the method detailed in section 2 after the $CO_2$ was completely desorbed. The measured porosity was assumed to be representative for $f_0$.

$$a_0 \approx \left(\frac{3f_0}{4\pi N_d}\right)^{\frac{1}{3}} \tag{29}$$

where the cell nucleation density $N_d$ equals $2\times 10^{20}$ m$^{-3}$ for the low $M_w$ PMMA nanofoams (see Table 1) and $N_d$ equals $20\times 10^{20}$ m$^{-3}$ for the high $M_w$ PMMA nanofoams (see Table 2).

## 5. Results and discussion of the void growth predictions

Consider the deformation mechanism maps for the low $M_w$ PMMA (see Figure 6a) and for the high $M_w$ PMMA (see Figure 6b). We superpose the predicted trajectory of the effective stress at the surface of the cavity $\sigma_e$ by the void growth model as a function of $T/T_g$ for foaming temperatures $T_f = 25$ °C and $T_f = 80$ °C, and for a foaming time up to 600 s. Note that both the temperature $T$ and glass transition temperature $T_g$ evolve in time during foaming. For both the low $M_w$ and high $M_w$ PMMA, at the start of foaming, $T$ equals $T_0$ and $T/T_g$ is close to 0.9; at this instant $\sigma_e$ is close to 0.8 MPa for the low $M_w$ PMMA and $\sigma_e$ is close to 0.3 MPa for the high $M_w$ PMMA. When the temperature increases from $T = T_0$ to $T = T_f$, $T/T_g$ rises to almost unity and $\sigma_e$ rises steeply. The void growth simulations suggest that during solid-state foaming of PMMA, the normalized temperature $T/T_g$ remains between 0.9 and 1 and consequently void growth does not occur within either the viscous regime (low $M_w$ PMMA) or within the rubbery regime (high $M_w$ PMMA).

The measured porosity $f$ is plotted as a function of foaming time $t_f$ for $T_f = 25$ °C to $T_f = 80$ °C, and compared with the predicted $f$ versus $t_f$ curves for the low $M_w$ and high $M_w$ nanofoams, in Figure 7a and Figure 7b, respectively. There is reasonably good agreement between the measured and the predicted $f$ - $t_f$ curves for $T_f = 25$ °C and $T_f = 40$ °C. The void growth model overestimates the porosity at $T_f = 60$ °C and at $T_f = 80$ °C, where porosities close to $f_{max}$ are observed. Observations of SEM micrographs suggest that cell walls tear, leading to open-celled microstructures. This is confirmed by open cell content measurements by gas pycnometry: nanofoams with the highest observed porosities have predominantly open-celled microstructures, see Figure 3. At increased foaming temperatures (i.e. $T_f = 100$ °C) collapse of the foamed open-celled microstructure is observed leading to measured porosities below the maximum observed porosities at $T_f = 80$ °C, as shown in Figures 2c and 2d.

We proceed to explore two alternative hypotheses for cell wall failure which could lead to open-celled microstructures as observed for the PMMA nanofoams: (i) achievement of a critical hoop strain at the void, or (ii) achievement of a minimum (critical) value of ligament thickness between neighbouring voids.

*(i) Critical hoop strain*

Assume that the solid surrounding the expanding void is incompressible. Then, by Eq. (3),

$$b^3 - a^3 = b_0^3 - a_0^3 \tag{30}$$

Recall that the initial (as-nucleated) porosity $f_0$ equals $(a_0/b_0)^3$ as defined in Eq. (28) and the current porosity $f$ equals $(a/b)^3$. Now, rearrange Eq. (30), to express $f$ as a function of $f_0$ and the true hoop strain $\varepsilon_s$ at the surface of the void, where $\varepsilon_s = \varepsilon_{\theta\theta}(r=a) = \ln(a/a_0)$:

$$f^{-1} = 1 + \exp(-3\varepsilon_s)(f_0^{-1} - 1) \tag{31}$$

Tearing of the cell wall occurs when $\varepsilon_s$ equals the $T/T_g$-dependent[20] true tensile failure strain $\varepsilon_f$. The critical porosity $f_f$ corresponding to this ductility-governed failure criterion reads:

$$f_f^{-1} = 1 + \exp(-3\varepsilon_f)(f_0^{-1} - 1) \tag{32}$$

*(ii) Critical ligament size*

The alternative failure hypothesis assumes that there is a minimum number of confined polymer chains separating individual cells to prevent rupture of the solid between the cells. Define the smallest distance between two neighbouring cells $h$ as:

$$h = 2(b - a) \tag{33}$$

---

[20] We assume $\varepsilon_f$ to be insensitive to strain rate [19,46].

Then, upon making use of the expressions $f_0 = (a_0/b_0)^3$, $f = (a/b)^3$, and Eq. (31), we obtain:

$$\frac{h}{a_0} = 2\left(f^{-\frac{1}{3}} - 1\right)\left(\frac{f_0^{-1} - 1}{f^{-1} - 1}\right)^{\frac{1}{3}} \qquad (34)$$

Write $h_c$ as the critical cell wall thickness, and assume that it is independent of the value of $T/T_g$. The corresponding critical value of porosity $f_c$ is given by Eq. (34) with $h = h_c$.

The ductility-governed porosity limit $f_f$ as given by Eq. (32) is plotted in Figure 7 based on the predicted hoop strain $\varepsilon_s$ during void growth. Note that we make use of the measured response of $\varepsilon_f$ versus $T/T_g$ (Eq. (A.2) for the low $M_w$ PMMA and Eq. (A.1) for the high $M_w$ PMMA as detailed in the Appendix) and assume that the initial porosity $f_0$ equals $10^{-3}$. The measured values of final porosity $f$ and the predictions of the void growth model exceed the porosity limit as given by $f_f$.

We now plot the porosity limit $f_c$ in Figure 7 via Eq. (34) for $f_0 = 10^{-3}$ by taking $h_c/a_0 = 3$ (low $M_w$ PMMA) and $h_c/a_0 = 4.2$ (high $M_w$ PMMA) in order to match to observed values of the maximum observed porosity $f_{max}$ of the nanofoams. Recall that the initial void size $a_0$ of the low $M_w$ PMMA nanofoams is estimated to be close to 10.5 nm, whereas $a_0$ is close to 5 nm for the high $M_w$ PMMA nanofoams. Consequently, the estimated corresponding critical cell wall dimension $h_c$ equals 32 nm for the low $M_w$ PMMA nanofoams, whereas $h_c$ equals 21 nm for the high $M_w$ PMMA. These values for $h_c$ are of the same order of magnitude as root-mean-square end-to-end distance $R_{ee}$ of the PMMA chains, i.e. $R_{ee} \approx 20$ nm for the low $M_w$ PMMA and $R_{ee} \approx 110$ nm for the high $M_w$ PMMA based on an idealized equivalent freely jointed chain calculation [43]. This is in agreement with the results of Crosby and co-workers who conducted a series of uniaxial tensile tests on thin polystyrene (PS) films with $M_w = 136\,000$ g mol$^{-1}$ [44,45]. They found that the tensile failure strain $\varepsilon_f$ decreases with decreasing film thickness $t$ in the regime $t = 15$ nm to $t = 77$ nm; these values are close to the estimated value for $R_{ee} = 25$ nm of the PS chains.

**Concluding remarks**

Solid-state nanofoaming experiments are performed with two grades of PMMA of markedly different molecular weight ($M_w$ = 92 500 g mol$^{-1}$ and $M_w$ = 3 580 000 g mol$^{-1}$). It was found that the molecular weight of the PMMA has a profound effect upon the microstructure of the PMMA nanofoams. When subjected to identical foaming conditions, the observed cell size $l \approx 35$ nm of the high molecular weight PMMA nanofoams is an order of magnitude less than that of the low molecular weight PMMA nanofoams, $l \approx 250$ nm. This is consistent with the observation that the nucleation density, $N_d \approx 20 \times 10^{20}$ m$^{-3}$ of the high molecular weight PMMA nanofoams is an order of magnitude higher than that of the low molecular weight PMMA nanofoams $N_d \approx 2 \times 10^{20}$ m$^{-3}$. In addition, a limit in attainable porosity $f_{max}$ was observed: $f_{max}$ equals 0.65 for the high molecular weight PMMA and $f_{max}$ equals 0.75 for the low molecular weight PMMA. The microstructure of the PMMA nanofoams transitions from closed-celled to open-celled at a porosity close to $f_{max}$.

A void growth model has been developed to simulate cavity expansion during solid-state nanofoaming of PMMA by $CO_2$. Experimentally calibrated constitutive laws for the PMMA grades close to the glass transition are used in the simulations. The predicted porosity of the nanofoams versus foaming time, at selected foaming temperatures, are in good agreement with the measured responses for porosities well below the maximum observed porosity. There is also close agreement between the predicted and observed sensitivity to molecular weight. This suggests that the observed difference in constitutive response close to the glass transition between the two PMMA grades leads to the measured difference in porosity. Moreover, cell wall tearing accounts for the observed limit in final porosity. Our analysis suggests the existence of a limiting minimum cell wall thickness of magnitude close to that of the end-to-end distance of the polymer chains. When the cell wall thickness approaches this minimum value during foaming, rupture of the cell walls occurs; resulting in an open-celled structure, and to a limit on foam expansion.

**Acknowledgements**

Financial support from the Engineering and Physical Sciences Research Council (UK) award 1611305 (F. Van Loock) and the FPU grant FPU14/02050 (V. Bernardo) from the Spanish Ministry of Education is gratefully acknowledged. N. A. Fleck is grateful for additional financial support from the ERC project MULTILAT. Financial assistance from SABIC,

Preprint of http://doi.org/10.1098/rspa.2019.0339

MINECO, FEDER, UE (MAT2015-69234-R) are acknowledged too. The authors would also like to thank Dr. Martin van Es from SABIC for the technical assistance.

**List of table captions**

Table 1: Measured porosity $f$, average cell size $l$, standard deviation of observed cell size $s$, cell nucleation density $N_\mathrm{d}$, and open cell content $O_\mathrm{v}$ of the low $M_\mathrm{w}$ PMMA nanofoams as a function of foaming time $t_\mathrm{f}$ and foaming temperature $T_\mathrm{f}$. Foams collapsed at $T_\mathrm{f} = 100\,°\mathrm{C}$, and so no open cell content values are reported for nanofoams produced at $T_\mathrm{f} = 100\,°\mathrm{C}$.

Table 2: Measured values for the porosity $f$, the average observed cell size $l$, the standard deviation of the observed cell sizes $s$, the cell nucleation density $N_\mathrm{d}$, and the open cell content $O_\mathrm{v}$ of the high $M_\mathrm{w}$ PMMA nanofoams as a function of foaming time $t_\mathrm{f}$ and foaming temperature $T_\mathrm{f}$. Foams collapsed at $T_\mathrm{f} = 100\,°\mathrm{C}$, and so no open cell content values are reported for the nanofoams produced at $T_\mathrm{f} = 100\,°\mathrm{C}$.

Table 3: Fitted parameters for the constitutive laws for the low $M_\mathrm{w}$ PMMA (Eq. (21) and Eq. (23)) and the high $M_\mathrm{w}$ PMMA obtained from Van Loock and Fleck [18], see Eq. (21) and Eq. (22).

Table 4: Summary of the assumed processing parameters and material properties for the void growth predictions.

**Tables**

Table 1: Measured porosity $f$, average cell size $l$, standard deviation of observed cell size $s$, cell nucleation density $N_d$, and open cell content $O_v$ of the low $M_w$ PMMA nanofoams as a function of foaming time $t_f$ and foaming temperature $T_f$. Foams collapsed at $T_f = 100\,°C$, and so no open cell content values are reported for nanofoams produced at $T_f = 100\,°C$.

| $t_f$ (s) | $T_f$ (°C) | $f$ | $l$ (nm) | $s$ (nm) | $N_d$ ($10^{20}$ m$^{-3}$) | $O_v$ |
|---|---|---|---|---|---|---|
| 60 | 25 | 0.45 | 219 | 87 | 1.50 | 0.12 |
| 180 | 25 | 0.47 | 228 | 79 | 1.50 | 0.08 |
| 300 | 25 | 0.51 | 283 | 112 | 0.91 | 0.08 |
| 600 | 25 | 0.51 | 235 | 85 | 1.48 | 0.08 |
| 60 | 40 | 0.52 | 262 | 102 | 1.22 | 0.07 |
| 180 | 40 | 0.61 | 250 | 125 | 1.70 | 0.02 |
| 300 | 40 | 0.64 | 254 | 105 | 1.27 | 0.15 |
| 600 | 40 | 0.66 | 233 | 103 | 2.11 | 0.14 |
| 60 | 60 | 0.56 | 234 | 89 | 2.34 | 0.07 |
| 180 | 60 | 0.66 | 297 | 111 | 1.72 | 0.33 |
| 300 | 60 | 0.68 | 279 | 122 | 1.76 | 0.40 |
| 600 | 60 | 0.68 | 284 | 109 | 1.63 | 0.36 |
| 60 | 80 | 0.72 | 333 | 134 | 1.16 | 0.63 |
| 180 | 80 | 0.74 | 288 | 138 | 1.83 | 0.90 |
| 300 | 80 | 0.75 | 297 | 125 | 1.75 | 0.78 |
| 600 | 80 | 0.73 | 274 | 109 | 2.08 | 0.93 |
| 60 | 100 | 0.64 | 297 | 122 | 1.21 | - |
| 180 | 100 | 0.68 | 253 | 110 | 1.81 | - |
| 300 | 100 | 0.62 | 246 | 103 | 1.75 | - |
| 600 | 100 | 0.51 | 291 | 125 | 0.76 | - |

Table 2: Measured values for the porosity $f$, the average observed cell size $l$, the standard deviation of the observed cell sizes $s$, the cell nucleation density $N_d$, and the open cell content $O_v$ of the high $M_w$ PMMA nanofoams as a function of foaming time $t_f$ and foaming temperature $T_f$. Foams collapsed at $T_f = 100\,°C$, and so no open cell content values are reported for the nanofoams produced at $T_f = 100\,°C$.

| $t_f$ (s) | $T_f$ (°C) | $f$ | $l$ (nm) | $s$ (nm) | $N_d$ ($10^{20}$ m$^{-3}$) | $O_v$ |
|---|---|---|---|---|---|---|
| 60 | 25 | 0.22 | 36 | 14 | 14.9 | 0.30 |
| 180 | 25 | 0.28 | 23 | 10 | 40.0 | 0.22 |
| 300 | 25 | 0.29 | 30 | 12 | 9.0 | 0.28 |
| 600 | 25 | 0.31 | 36 | 18 | 6.9 | 0.21 |
| 60 | 40 | 0.33 | 28 | 13 | 54.2 | 0.19 |
| 180 | 40 | 0.42 | 32 | 16 | 32.3 | 0.07 |
| 300 | 40 | 0.45 | 37 | 14 | 7.8 | 0.08 |
| 600 | 40 | 0.47 | 45 | 29 | 26.0 | 0.09 |
| 60 | 60 | 0.45 | 37 | 14 | 20.4 | 0.08 |
| 180 | 60 | 0.55 | 39 | 17 | 24.0 | 0.03 |
| 300 | 60 | 0.57 | 40 | 17 | 31.8 | 0.28 |
| 600 | 60 | 0.57 | 41 | 19 | 25.8 | 0.03 |
| 60 | 80 | 0.58 | 39 | 20 | 21.8 | 0.51 |
| 180 | 80 | 0.60 | 39 | 19 | 27.8 | 0.73 |
| 300 | 80 | 0.60 | 38 | 19 | 36.6 | 0.95 |
| 600 | 80 | 0.59 | 44 | 22 | 46.6 | 0.88 |
| 60 | 100 | 0.59 | 34 | 15 | 35.4 | - |
| 180 | 100 | 0.53 | 27 | 14 | 80.4 | - |
| 300 | 100 | 0.50 | 37 | 18 | 24.9 | - |
| 600 | 100 | 0.45 | 34 | 12 | 32.6 | - |

Table 3: Fitted parameters for the constitutive laws for the low $M_w$ PMMA (Eq. (21) and Eq. (23)) and the high $M_w$ PMMA obtained from Van Loock and Fleck [19], see Eq. (21) and Eq. (22).

|  | low $M_w$ PMMA | high $M_w$ PMMA |
|---|---|---|
| $v$ (nm$^{-3}$) | 2.5 | 1.8 |
| $q$ (J) | $7.31 \times 10^{-19}$ | $7.31 \times 10^{-19}$ |
| $\dot{\varepsilon}_0$ (s$^{-1}$) | $1.5 \times 10^{56}$ | $1.5 \times 10^{56}$ |
| $\eta_0$ (Pa s) | $2.8 \times 10^6$ | - |
| $C_1$ | 3.2 | - |
| $C_2$ (K) | 17.3 | - |
| $E_R^0$ (MPa) | - | 65.8 |
| $\alpha_R$ | - | 0.8 |
| $\dot{\varepsilon}_R$ (s$^{-1}$) | - | 1.58 |
| $n$ | - | 0.173 |

Table 4: Summary of the assumed processing parameters and material properties for the void growth predictions.

|  | low $M_w$ PMMA | high $M_w$ PMMA |
|---|---|---|
| $p_0$ (MPa) | 31 | 31 |
| $p_a$ (MPa) | 0.1 | 0.1 |
| $\tau$ (s) | 20 | 20 |
| $\rho^p$ (kg m$^{-3}$) | 1190 | 1190 |
| $T_g$ (°C) | 114.5 | 116.5 |
| $f_0$ | $10^{-3}$ | $10^{-3}$ |
| $a_0$ (nm) | 10.5 | 5 |

**List of Figure captions**

Figure 1: SEM micrographs of the low $M_w$ nanofoams at (a) $T_f = 60\ °C$, (b) $T_f = 100\ °C$ and of the high $M_w$ nanofoams at (c) $T_f = 60\ °C$ and (d) $T_f = 100\ °C$.

Figure 2: Nanofoaming experiments on the low $M_w$ and high $M_w$ PMMA grades: (a) measured average cell size $l$ versus foaming time $t_f$ for $T_f = 60\ °C$, (b) measured porosity $f$ versus foaming time $t_f$ for $T_f = 60\ °C$ and $T_f = 100\ °C$, (c) measured porosity $f$ versus foaming temperature $T_f$ for the range of explored foaming times ($t_f = 60$ s to $t_f = 600$ s) for the low $M_w$ nanofoams, and (d) measured $f$ versus $T_f$ for the range of explored foaming times ($t_f = 60$ s to $t_f = 600$ s) for the high $M_w$ nanofoams.

Figure 3: Measured open cell content $O_v$ as a function of porosity $f$ for (a) the low $M_w$ PMMA nanofoam and (b) high $M_w$ PMMA nanofoam.

Figure 4: Spherical void in (a) undeformed configuration with initial radius $a_0$ and initial outer radius $b_0$ and (b) deformed configuration at time $t$ of the void with radius $a$, outer radius $b$ and gas pressure $p$.

Figure 5: The normalized glass transition temperature $T_g / T_g^0$ of PMMA as a function of $CO_2$ mass concentration $C$, as reported by Chiou *et al.* [30], Wissinger and Paulaitis [31], and Guo and Kumar [26]. The $T_g / T_g^0$ versus $C$ curve is given by the calibrated version of Eq. (20).

Figure 6: Deformation mechanism maps for (a) low $M_w$ PMMA and (b) high $M_w$ PMMA (for a reference strain $\varepsilon_{ref} = 0.05$), for contours of effective strain rate $\dot\varepsilon_e$. The predicted effective stress at the surface of the cavity $\sigma_e$ is plotted as a function of $T/T_g$ for foaming temperatures $T_f = 25\ °C$ and $T_f = 80\ °C$ and for a foaming time up to 600 s.

Figure 7: Predicted and measured porosity $f$ versus foaming time $t_f$, for $T_f = 25\ °C$ to $T_f = 80\ °C$ for (a) low $M_w$ nanofoams and (b) high $M_w$ PMMA nanofoams. The ductility-governed porosity limit $f_f$ is plotted via Eq. (32) for an initial porosity $f_0 = 10^{-3}$. The minimum cell wall thickness-governed porosity limit $f_c$ is plotted via Eq. (34) for $f_0 = 10^{-3}$ and $h_c/a_0 = 3$ (low $M_w$ PMMA) and $h_c/a_0 = 4.2$ (high $M_w$ PMMA).

**Figures**

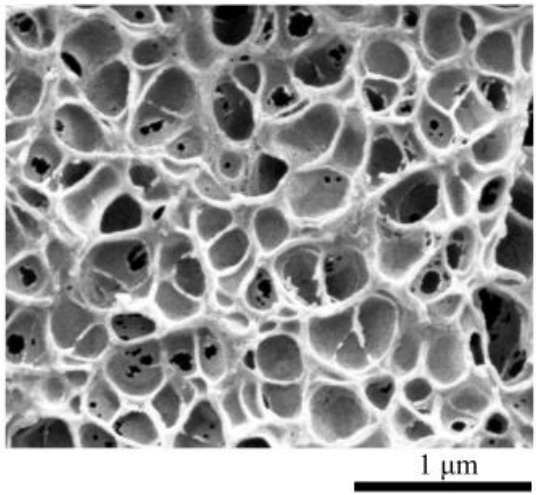
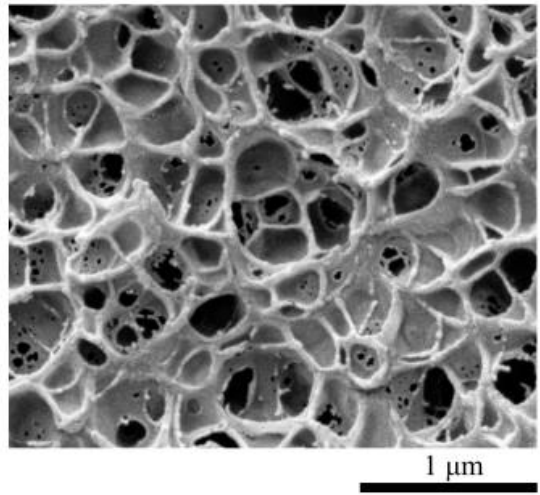
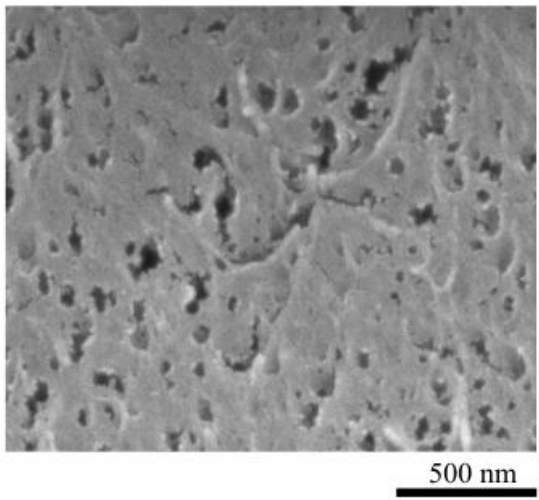
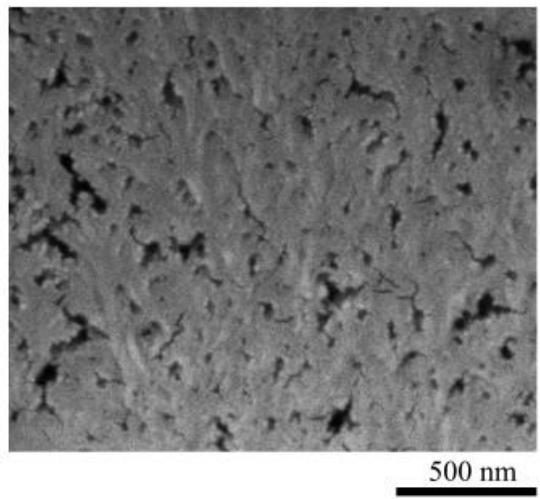

Figure 1: SEM micrographs of the low $M_w$ nanofoams at (a) $T_f = 60\ °C$, (b) $T_f = 100\ °C$ and of the high $M_w$ nanofoams at (c) $T_f = 60\ °C$ and (d) $T_f = 100\ °C$.

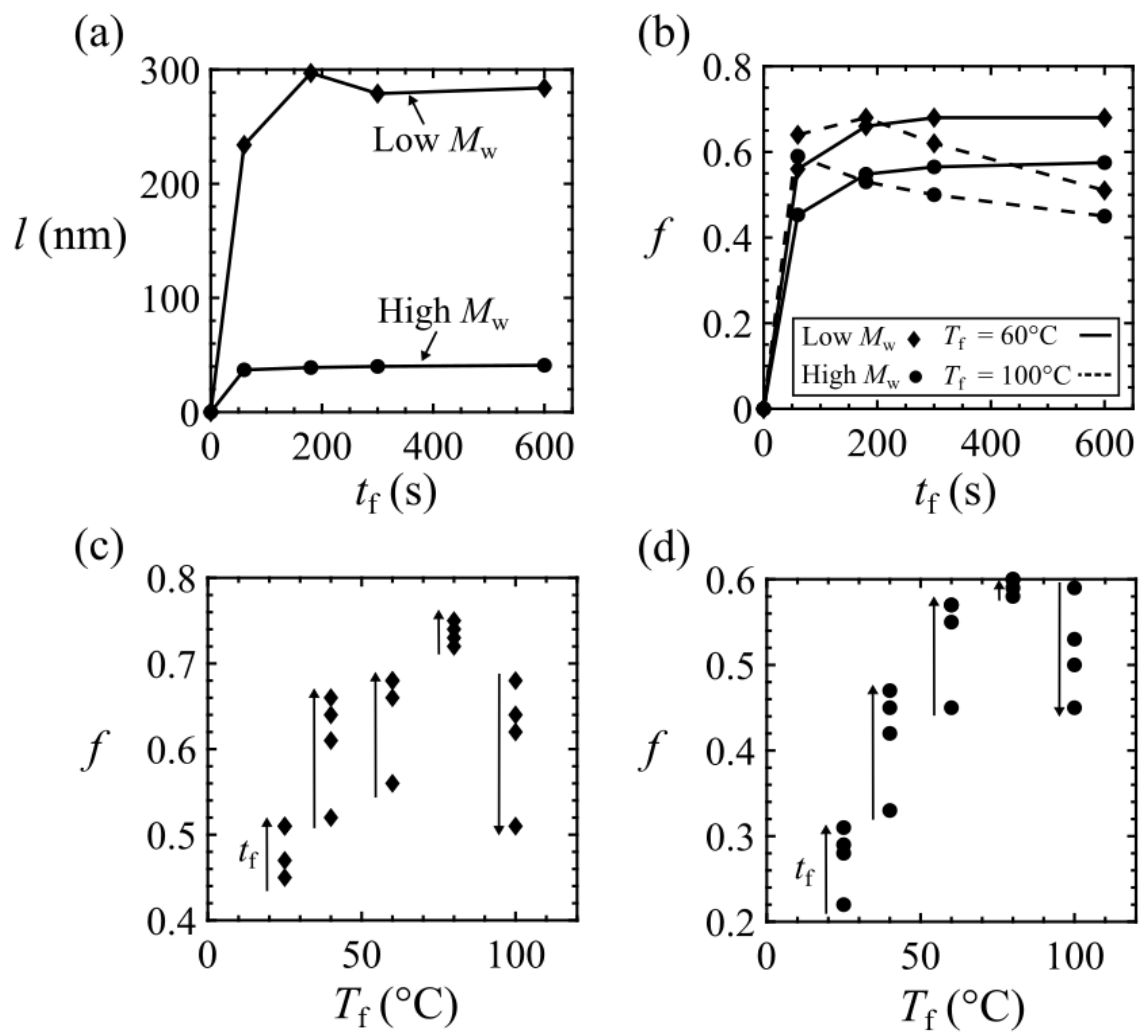

Figure 2: Nanofoaming experiments on the low $M_w$ and high $M_w$ PMMA grades: (a) measured average cell size $l$ versus foaming time $t_f$ for $T_f = 60$ °C, (b) measured porosity $f$ versus foaming time $t_f$ for $T_f = 60$ °C and $T_f = 100$ °C, (c) measured porosity $f$ versus foaming temperature $T_f$ for the range of explored foaming times ($t_f = 60$ s to $t_f = 600$ s) for the low $M_w$ nanofoams, and (d) measured $f$ versus $T_f$ for the range of explored foaming times ($t_f = 60$ s to $t_f = 600$ s) for the high $M_w$ nanofoams.

Preprint of http://doi.org/10.1098/rspa.2019.0339

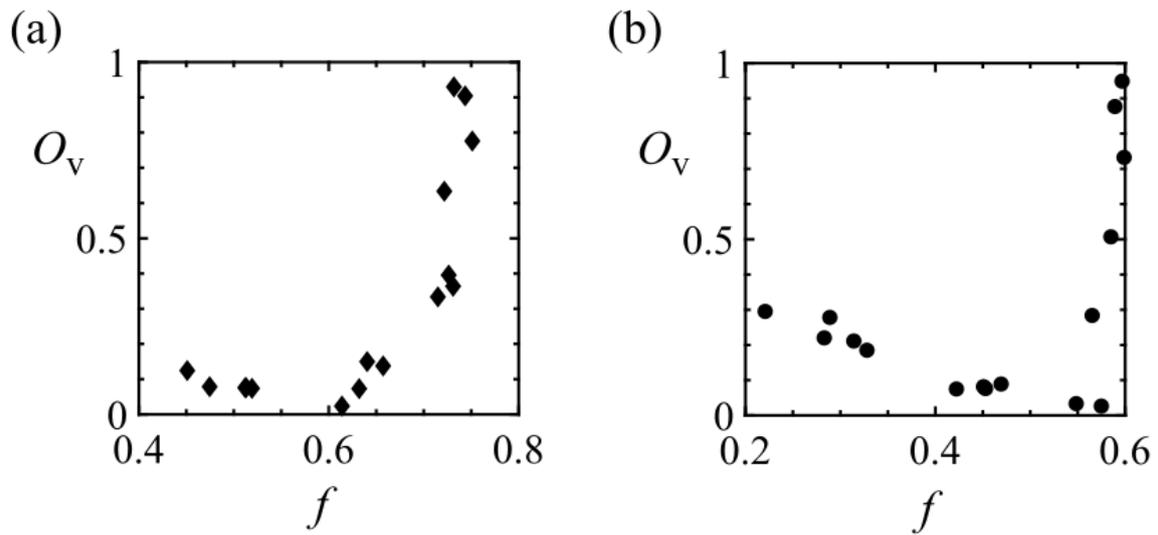

Figure 3: Measured open cell content $O_v$ as a function of porosity $f$ for (a) the low $M_w$ PMMA nanofoam and (b) high $M_w$ PMMA nanofoam.

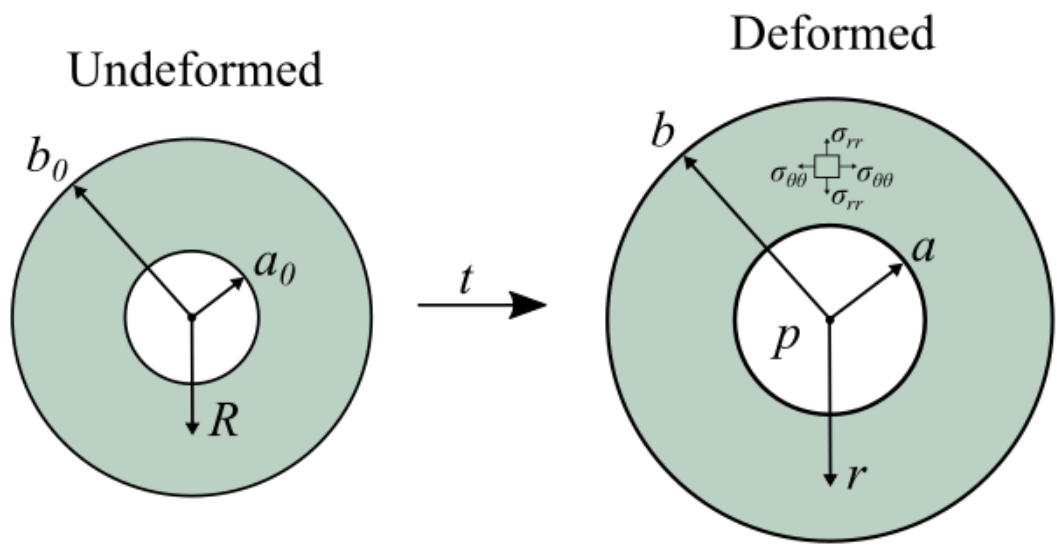

Figure 3: Spherical void in (a) undeformed configuration with initial radius $a_0$ and initial outer radius $b_0$ and (b) deformed configuration at time $t$ of the void with radius $a$, outer radius $b$ and gas pressure $p$.

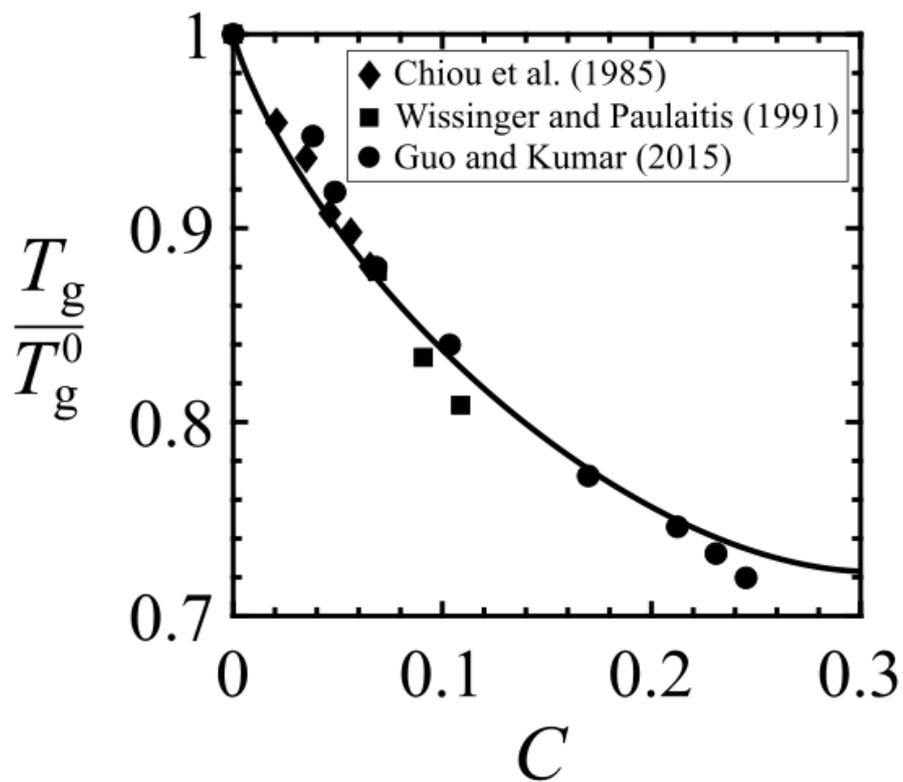

Figure 5: The normalized glass transition temperature $T_g/T_g^0$ of PMMA as a function of $CO_2$ mass concentration $C$, as reported by Chiou *et al.* [32], Wissinger and Paulaitis [33], and Guo and Kumar [28]. The $T_g/T_g^0$ versus $C$ curve is given by the calibrated version of Eq. (20).

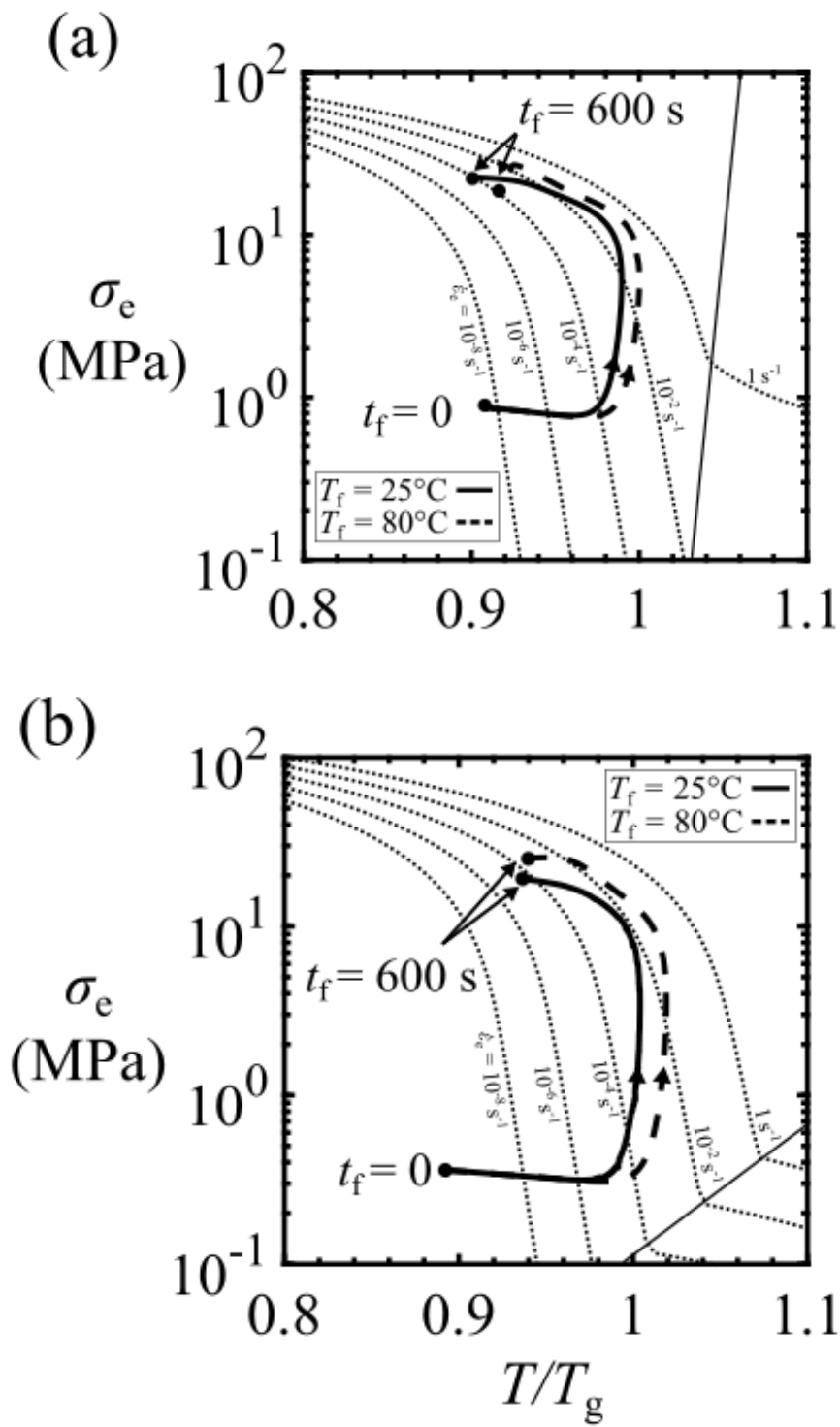

Figure 6: Deformation mechanism maps for (a) low $M_w$ PMMA and (b) high $M_w$ PMMA (for a reference strain $\varepsilon_{ref} = 0.05$), for contours of effective strain rate $\dot{\varepsilon}_e$. The predicted effective stress at the surface of the cavity $\sigma_e$ is plotted as a function of $T/T_g$ for foaming temperatures $T_f = 25$ °C and $T_f = 80$ °C and for a foaming time up to 600 s.

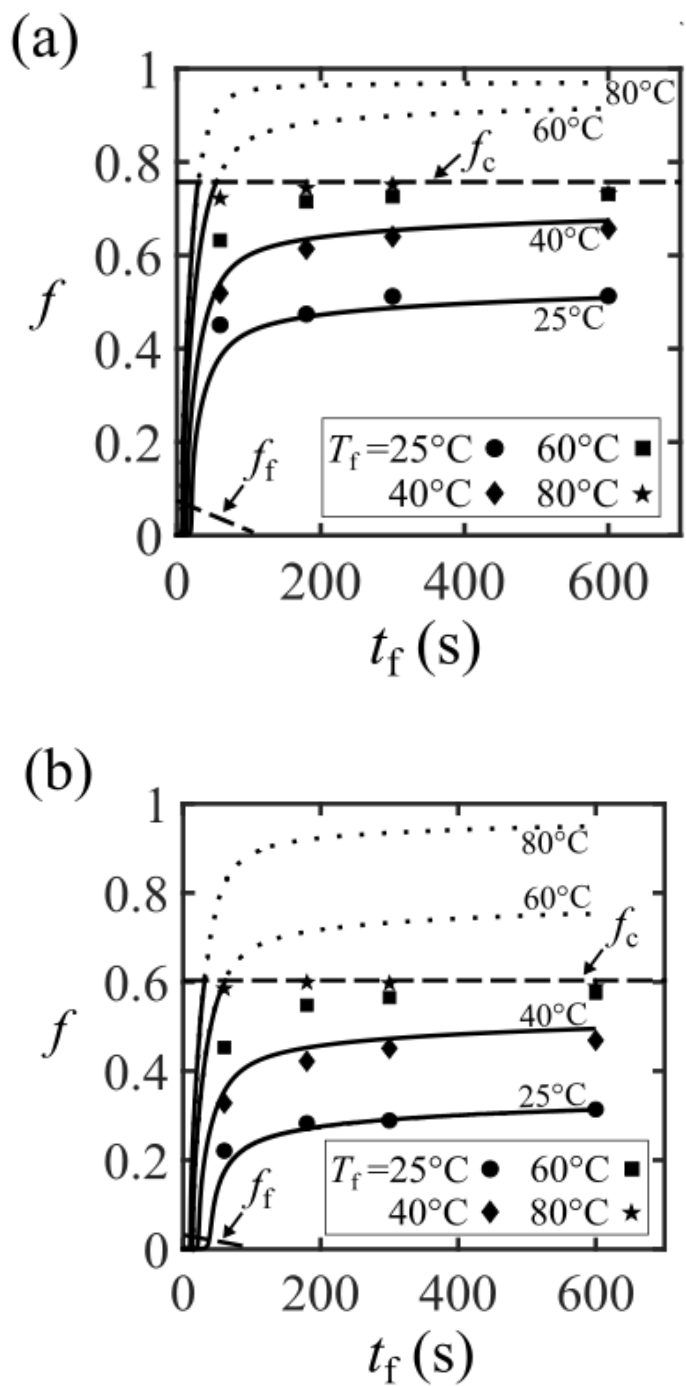

Figure 4: Predicted and measured porosity $f$ versus foaming time $t_f$, for $T_f = 25\,°C$ to $T_f = 80\,°C$ for (a) low $M_w$ nanofoams and (b) high $M_w$ PMMA nanofoams. The ductility-governed porosity limit $f_f$ is plotted via Eq. (32) for an initial porosity $f_o = 10^{-3}$. The minimum cell wall thickness-governed porosity limit $f_c$ is plotted via Eq. (34) for $f_o = 10^{-3}$ and $h_c/a_0 = 3$ (low $M_w$ PMMA) and $h_c/a_0 = 4.2$ (high $M_w$ PMMA).

**Appendix**

*Calibration of constitutive laws for the low $M_w$ and high $M_w$ PMMA*

Constitutive laws are calibrated for the low $M_w$ PMMA grade[21] close to its glass transition temperature. We follow the procedure of Van Loock and Fleck [19] who constructed deformation and failure maps for the high $M_w$ PMMA grade[22] in uniaxial tension close to the glass transition temperature. A series of uniaxial tensile tests were performed on the low $M_w$ PMMA grade for a range of temperatures (T = 90 °C to T = 170 °C) and at a nominal strain rate $\dot{e} = 5.9 \times 10^{-2}$ s$^{-1}$. The dogbone specimen geometry and the measurement procedures are detailed in Van Loock and Fleck [19]. Note that the low $M_w$ PMMA dogbone specimens are machined from the foaming precursor sheets. The true stress versus true strain responses of the low $M_w$ PMMA dogbone specimens are plotted in Figure A1.a for $0.94 < T/T_g < 1.01$ and in Figure A.1b $1.04 < T/T_g < 1.14$. The true stress versus true strain curves of the high $M_w$ PMMA grade are included in Figures A.1a and A.1b.

Loading-unloading uniaxial stress stress versus strain curves for the low $M_w$ PMMA and high low $M_w$ PMM are shown in Figure A.2. At $T/T_g = 0.93$, the elastic unloading of the low $M_w$ and the high $M_w$ PMMA occurs in the manner of an elasto-viscoplastic solid, with remnant a finite strain at zero load. The qualitative stress versus strain response of the low $M_w$ and the high $M_w$ PMMA is different when the temperature is increased to $T/T_g = 1.06$. The elastic rubbery regime is entered for the high $M_w$ PMMA and the unloading curve is almost coincidental with the loading curve; there is negligible hysteresis and negligible remnant strain. No rubbery regime is observed for the low $M_w$ PMMA above the glass transition. At $T/T_g = 1.06$ and $T/T_g = 1.12$, the stress versus strain response of the low $M_w$ PMMA in uniaxial tension is linear viscous. Unloading is accompanied by a finite remnant strain. The high $M_w$ PMMA transitions from the rubbery regime to a viscous regime at $T/T_g = 1.16$.

---

[21] Altuglas V825T with $T_g = 114.5$ °C and $M_w = 92\ 500$ g mol$^{-1}$.

[22] Altuglas CN with $T_g = 116.5$ °C and $M_w = 3\ 580\ 000$ g mol$^{-1}$.

First, consider the elasto-viscoplastic regime. The dependence of the measured flow strength $\sigma_y$ of the low $M_w$ and high $M_w$ PMMA grades upon $T/T_g$ is shown in Figure A.3 for $\dot{e} = 5.9 \times 10^{-2}$ s$^{-1}$. A single transition Ree-Eyring equation, Eq. (21), is fitted to the $\sigma_y$ versus $T/T_g$ response of the low $M_w$ PMMA in the glassy and glass transition regime (corresponding to $0.94 \leq T/T_g \leq 1.04$). We assume that $q$ equals $7.31 \times 10^{-19}$ J and $\dot{\varepsilon}_0$ equals $= 1.5 \times 10^{56}$ s$^{-1}$ for both the low $M_w$ and the high $M_w$ PMMA, as reported by Van Loock and Fleck [19]. The activation volume $v$ equals 2.5 nm$^{-3}$ for the low $M_w$ PMMA, and $v = 1.8$ nm$^{-3}$ for the high $M_w$ PMMA [19]. The resulting curve fits are included in Figure A.3. Second, consider the viscous regime for the low $M_w$ PMMA. We fit a linear, viscous constitutive law, Eqs. (23) and (24), to the measured $\sigma_y$ versus $T/T_g$ curves of the low $M_w$ PMMA in the regime of $1.06 \leq T/T_g \leq 1.14$ and $\dot{e} = 5.9 \times 10^{-2}$ s$^{-1}$. The fitting values are $\eta_0 = 2.8 \times 10^6$ Pa·s, $C_1 = 3.2$, and $C_2 = 17.3$ K. The resulting curve fit is adequate, see Figure A.3. Third, consider the rubbery regime of the high $M_w$ PMMA. The constitutive description, Eq. (22), is adequate upon making use of previously measured values ($E_R^0 = 65.8$ MPa, $\alpha_R = 0.80$, $\dot{\varepsilon}_R = 1.58$ s$^{-1}$, and $n = 0.173$ [19]), as shown in Figure A.3.

*Tensile ductility of the low $M_w$ and high $M_w$ PMMA*

Van Loock and Fleck [19] measured the true tensile failure strain, that is ductility, $\varepsilon_f$ of the high $M_w$ PMMA grade by testing a dogbone geometry at $T/T_g < 1$ and an hourglass-shaped specimen geometry at $T/T_g \geq 1$. The measured values for $\varepsilon_f$ of the high $M_w$ PMMA grade are plotted as a function of the normalized temperature $T/T_g$ for a nominal strain rate $\dot{e} = 5.9 \times 10^{-2}$ s$^{-1}$ in Figure A.4. The $\varepsilon_f$ versus $T/T_g$ failure envelope is adequately fitted by a linear relation [19]:

$$\varepsilon_f = 7.3 \frac{T}{T_g} - 6.3 \tag{A.1}$$

An additional series of uniaxial tensile tests have been conducted on the low $M_w$ PMMA grade by using the same measurement methods as that detailed in the work of Van Loock and Fleck [19]. No failure was observed at $T \geq 145\,°C$ prior to the attainment of the maximum cross-head extension. The measured $\varepsilon_f$ versus $T/T_g$ curve is shown in Figure A.4. The failure envelope of the low $M_w$ PMMA grade close to the glass transition is also fitted by a linear relation:

$$\varepsilon_f = 13.3 \frac{T}{T_g} - 11.7 \tag{A.2}$$

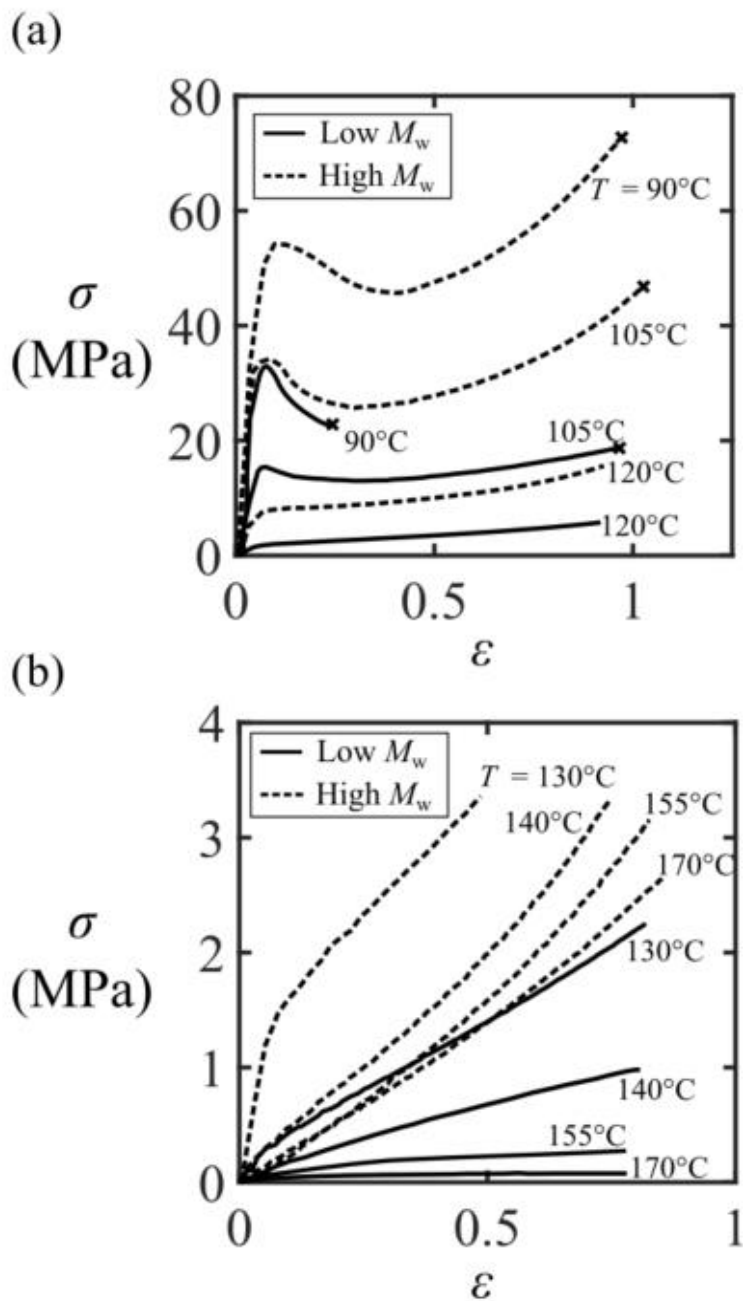

Figure A.1: Measured true tensile stress $\sigma$ versus true tensile strain $\varepsilon$ curves for the low $M_w$ and high $M_w$ PMMA grades in uniaxial tension for a nominal strain rate $\dot{\varepsilon} = 5.9 \times 10^{-2} \text{s}^{-1}$ and for temperatures ranging from (a) T = 90 °C to T = 120 °C and (b) T = 130 °C to T = 170 °C. A cross at the end of the curve denotes specimen failure.

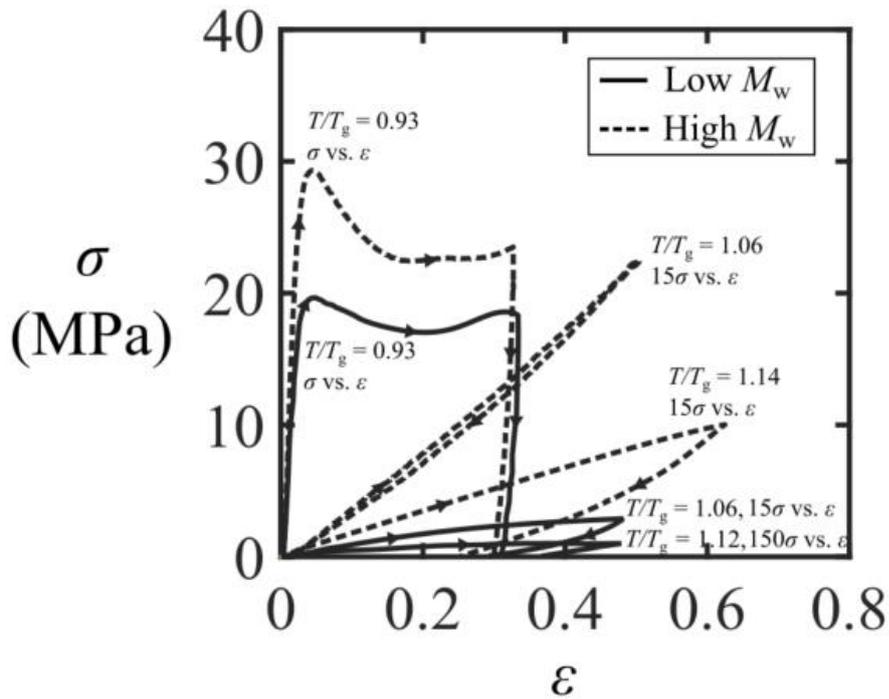

Figure A.2: Loading-unloading true stress versus true strain curves for the low $M_w$ PMMA and high $M_w$ PMMA grades in uniaxial tension, at selected values of $T/T_g$, for a nominal strain rate $\dot{\varepsilon} = 5.9 \times 10^{-4}\,\mathrm{s}^{-1}$.

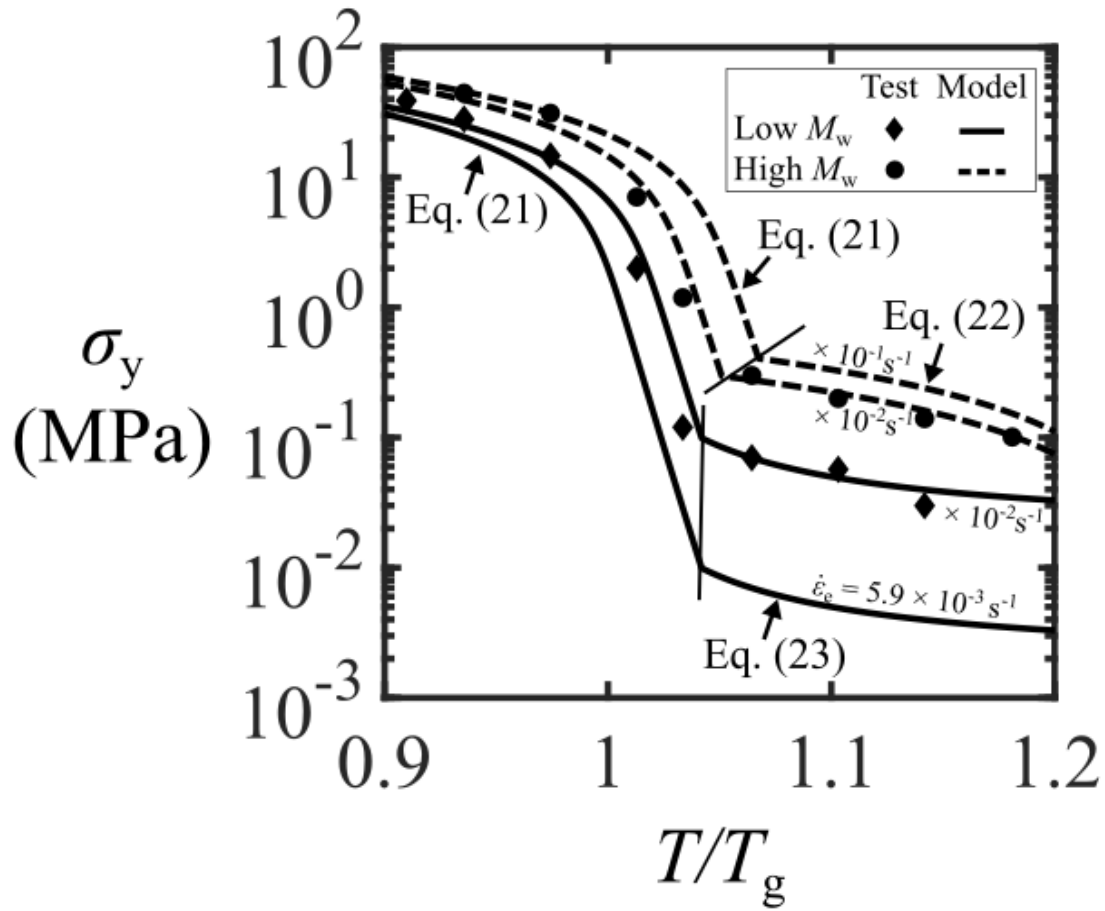

Figure A.3: Deformation mechanism maps of the low $M_w$ and high $M_w$ PMMA grades. Flow strength $\sigma_y$ ($= \sigma_e$) versus $T/T_g$ is plotted, with the curve fits of the constitutive models included for a reference strain $\varepsilon_{ref} = 0.05$.

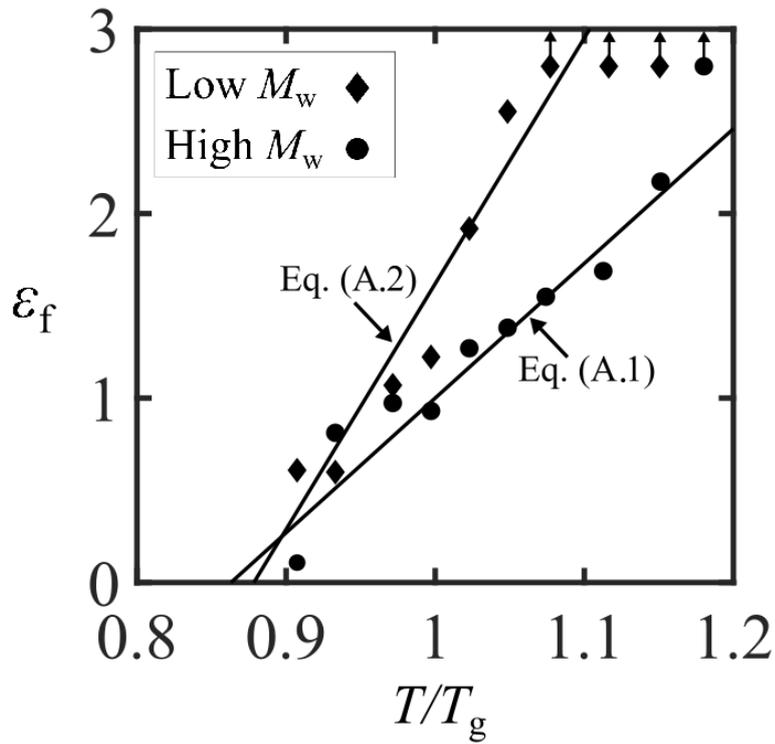

Figure A.4: The measured true tensile failure strain $\varepsilon_f$ as a function of normalized testing temperature $T/T_g$ for the low $M_w$ and high $M_w$ PMMA grades, at a nominal strain rate $\dot{\varepsilon}_e = 5.9 \times 10^{-2}\,\mathrm{s}^{-1}$.

Preprint of http://doi.org/10.1098/rspa.2019.0339

**References**

1. Costeux S. 2014 CO2-blown nanocellular foams. *J. Appl. Polym. Sci.* **131**.

2. Notario B, Pinto J, Rodríguez-Pérez MÁ. 2015 Towards a new generation of polymeric foams: PMMA nanocellular foams with enhanced physical properties. *Polymer (Guildf).* **63**, 116–126. (doi:10.1016/j.polymer.2015.03.003)

3. Miller D, Kumar V. 2009 Microcellular and nanocellular solid-state polyetherimide (PEI) foams using sub-critical carbon dioxide I. Processing and structure. *Polymer (Guildf).* **50**, 5576–5584. (doi:10.1016/j.polymer.2011.04.049)

4. Martín-de León J, Bernardo V, Rodríguez-Pérez MÁ. 2017 Key production parameters to obtain transparent nanocellular PMMA. *Macromol. Mater. Eng.* **302**, 3–7. (doi:10.1002/mame.201700343)

5. Schmidt D, Raman VI, Egger C, du Fresne C, Schadler V. 2007 Templated cross-linking reactions for designing nanoporous materials. *Mater. Sci. Eng. C* **27**, 1487–1490. (doi:10.1016/j.msec.2006.06.028)

6. Notario B, Pinto J, Solorzano E, de Saja JA, Dumon M, Rodríguez-Pérez MÁ. 2014 Experimental validation of the Knudsen effect in nanocellular polymeric foams. *Polymer (Guildf).* **56**, 57–67.

7. Wang G, Wang C, Zhao J, Wang G, Park CB, Zhao G. 2017 Modelling of thermal transport through a nanocellular polymer foam: toward the generation of a new superinsulating material. *Nanoscale* **9**, 5996–6009.

8. Martini JE. 1981 The production and analysis of microcellular foam (PhD thesis). MIT (USA).

9. Martin-de León J, Bernardo V, Rodríguez-Pérez MÁ. 2016 Low density nanocellular polymers based on PMMA produced by gas dissolution foaming : fabrication and cellular structure characterization. *Polymers (Basel).* **8**, 256.

10. Aher B, Olson NM, Kumar V. 2013 Production of bulk solid-state PEI nanofoams using supercritical CO2. *J. Mater. Res.* **28**, 2366–2373. (doi:10.1557/jmr.2013.108)

11. Costeux S, Zhu L. 2013 Low density thermoplastic nanofoams nucleated by nanoparticles. *Polymer (Guildf).* **54**, 2785–2795. (doi:10.1016/j.polymer.2013.03.052)

12. Bernardo V, Martin-de Leon J, Pinto J, Catelani T, Athanassiou A, Rodriguez-Perez MA. 2019 Low-density PMMA/MAM nanocellular polymers using low MAM contents: Production and characterization. *Polymer (Guildf).* **163**, 115–124. (doi:10.1016/j.polymer.2018.12.057)

13. Costeux S, Khan I, Bunker SP, Jeon HK. 2014 Experimental study and modeling of nanofoams formation from single phase acrylic copolymers. *J. Cell. Plast.* **51**, 197–221. (doi:10.1177/0021955X14531972)

14. Street JR. 1968 The Rheology of Phase Growth in Elastic Liquids. *J. Rheol. (N. Y. N. Y).* **12**, 103. (doi:10.1122/1.549101)

15. Amon M, Denson CD. 1984 A study of the dynamics of foam growth: Analysis of the

growth of closely spaced spherical bubbles. *Polym. Eng. Sci.* **24**, 1026–1034. (doi:10.1002/pen.760241306)

16. Venerus DC, Yala N, Bernstein B. 1998 Analysis of diffusion-induced bubble growth in viscoelastic liquids. *J. Nonnewton. Fluid Mech.* **75**, 55–75. (doi:10.1016/S0377-0257(97)00076-1)

17. Khan I, Adrian D, Costeux S. 2015 A model to predict the cell density and cell size distribution in nano-cellular foams. *Chem. Eng. Sci.* **138**, 634–645. (doi:10.1016/j.ces.2015.08.029)

18. Shafi MA, Lee JG, Flumerfelt RW. 1996 Prediction of cellular structure in free expansion polymer foam processing. *Polym. Eng. Sci.* **36**, 1950–1959. (doi:10.1002/pen.10591)

19. Van Loock F, Fleck NA. 2018 Deformation and failure maps for PMMA in uniaxial tension. *Polymer (Guildf).* **148**, 259–268. (doi:10.1016/j.polymer.2018.06.027)

20. Pinto J, Solórzano E, Rodriguez-Perez MA, De Saja JA. 2013 Characterization of the cellular structure based on user-interactive image analysis procedures. *J. Cell. Plast.* **49**, 555–575. (doi:10.1177/0021955X13503847)

21. Kumar V, Suh NP. 1990 A process for making microcellular thermoplastic parts. *Polym. Eng. Sci.* **30**, 1323–1329. (doi:10.1002/pen.760302010)

22. Rides M, Morikawa J, Halldahl L, Hay B, Lobo H, Dawson A, Allen C. 2009 Intercomparison of thermal conductivity and thermal diffusivity methods for plastics. *Polym. Test.* **28**, 480–489. (doi:10.1016/j.polymertesting.2009.03.002)

23. ASTM Int. 2015 Standard test method for open cell content of rigid cellular plastics.

24. Goel SK, Beckman EJ. 1995 Nucleation and Growth in Microcellular materials: Supercritical CO2 as foaming agent. *AIChe* **41**, 357–367. (doi:10.1002/aic.690410217)

25. Timoshenko SP, Goodier JN. 2001 *Theory of Elasticity*. McGraw-Hill.

26. Hill R. 1950 *The Mathematical Theory of Plasticity*. Oxford University Press Ic., New York.

27. Crank J. 1979 *The Mathematics of Diffusion*. Clarendon Press.

28. Guo H, Kumar V. 2015 Solid-state poly(methyl methacrylate) (PMMA) nanofoams. Part I: Low-temperature CO2 sorption, diffusion, and the depression in PMMA glass transition. *Polymer (Guildf).* **57**, 157–163.

29. Li R, Ye N, Shaayegan V, Fang T. 2018 Experimental measurement of CO2 diffusion in PMMA and its effect on microcellular foaming. *J. Supercrit. Fluids* **135**, 180–187. (doi:10.1016/j.supflu.2018.01.024)

30. Alessi P, Cortesi A, Kikic I, Vecchione F. 2003 Plasticization of polymers with supercritical carbon dioxide: Experimental determination of glass-transition temperatures. *J. Appl. Polym. Sci.* **88**, 2189–2193. (doi:10.1002/app.11881)

31. Verreck G, Decorte A, Li H, Tomasko D, Arien A, Peeters J, Rombaut P, Van den

Mooter G, Brewster ME. 2006 The effect of pressurized carbon dioxide as a plasticizer and foaming agent on the hot melt extrusion process and extrudate properties of pharmaceutical polymers. *J. Supercrit. Fluids* **38**, 383–391. (doi:10.1016/j.supflu.2005.11.022)

32. Chiou JS, Barlow JW, Paul DR. 1985 Plasticization of glassy polymers by CO2. *J. Appl. Polym. Sci.* **30**, 2633–2642. (doi:10.1002/app.1985.070300626)

33. Wissinger RG, Paulaitis ME. 1991 Glass transitions in polymer/CO2 mixtures at elevated pressures. *J. Polym. Sci. Part B Polym. Phys.* **29**, 631–633. (doi:10.1002/polb.1991.090290513)

34. Chow S. 1980 Molecular interpretation of the glass transition temperature of polymer-diluent systems. *Macromolecules* **364**, 362–364.

35. Li R, Li L, Zeng D, Liu Q, Fang T. 2016 Numerical Selection of the Parameters in Producing Microcellular Polymethyl Methacrylate With Supercritical CO 2. **35**, 309–328.

36. McLoughlin JR, Tobolsky A V. 1952 The viscoelastic behavior of polymethyl methacrylate. *J. Colloid Sci.* **7**, 555–568.

37. Williams ML, Landel RF, Ferry JD. 1955 The temperature dependence of relaxation mechanisms in amorphous polymers and other glass-forming liquids. *J. Am. Chem. Soc.* **77**, 3701–3707.

38. Bin Ahmad Z, Ashby MF. 1988 Failure-mechanism maps for engineering polymers. *J. Mater. Sci.* **23**, 2037–2050.

39. Rajendran A, Bonavoglia B, Forrer N, Storti G, Mazzotti M, Morbidelli M. 2005 Simultaneous measurement of swelling and sorption in a supercritical CO2-poly(methyl methacrylate) system. *Ind. Eng. Chem. Res.* **44**, 2549–2560. (doi:10.1021/ie049523w)

40. Pantoula M, Panayiotou C. 2006 Sorption and swelling in glassy polymer/carbon dioxide systems. *J. Supercrit. Fluids* **37**, 254–262. (doi:10.1016/j.supflu.2005.11.001)

41. Pantoula M, von Schnitzler J, Eggers R, Panayiotou C. 2007 Sorption and swelling in glassy polymer/carbon dioxide systems. Part II-Swelling. *J. Supercrit. Fluids* **39**, 426–434. (doi:10.1016/j.supflu.2006.03.010)

42. Van Krevelen DW, Te Nijenhuis K. 2009 *Properties of Polymers*. (doi:10.1016/B978-0-08-054819-7.X0001-5)

43. Rubinstein M, Colby HR. 2003 *Polymer Physics*. Oxford University Press.

44. Liu Y, Chen YC, Hutchens S, Lawrence J, Emrick T, Crosby AJ. 2015 Directly measuring the complete stress-strain response of ultrathin polymer films. *Macromolecules* **48**, 6534–6540. (doi:10.1021/acs.macromol.5b01473)

45. Bay RK, Shimomura S, Liu Y, Ilton M, Crosby AJ. 2018 Confinement effect on strain localizations in glassy polymer films. *Macromolecules* **51**, 3647–3653. (doi:10.1021/acs.macromol.8b00385)

46. Cheng WM, Miller GA, Manson JA, Hertzberg RW, Sperling LH. 1990 Mechanical behaviour of poly(methyl methacrylate). *J. Mater. Sci.* **25**, 1931–1938.

**Electronic supplied material for Figure 1 in publication version**

Martín-de León J, Bernardo V, Rodríguez-Pérez MÁ. 2017 Key production parameters to obtain transparent nanocellular PMMA. Macromol. Mater. Eng. 302, 3–7. (doi:10.1002/mame.201700343)

2,11

Guo H, Nicolae A, Kumar V. 2015 Solid-state poly(methyl methacrylate) (PMMA) nanofoams. Part II: Low-temperature solid-state process space using CO2 and the resulting morphologies. Polymer (Guildf). 70, 231–241. (doi:10.1016/j.polymer.2015.06.031)

Martín-de León J, Bernardo V, Rodríguez-Pérez MÁ. 2019 Nanocellular polymers: the challenge of creating cells in the nanoscale. Materials (Basel). 12, 797. (doi:10.3390/ma12050797)

4,7,8

Bernardo V, Martin-de Leon J, Pinto J, Catelani T, Athanassiou A, Rodriguez-Perez MA. 2019 Low-density PMMA/MAM nanocellular polymers using low MAM contents: Production and characterization. Polymer (Guildf). 163, 115–124. (doi:10.1016/j.polymer.2018.12.057)

5,6

Martin-de León J, Bernardo V, Rodríguez-Pérez MÁ. 2016 Low density nanocellular polymers based on PMMA produced by gas dissolution foaming : fabrication and cellular structure characterization. Polymers (Basel). 8, 256.

Bernardo V, Van Loock F, Martin-de Leon J, Fleck NA, Rodriguez-Perez MA. 2019 Mechanical Properties of PMMA-Sepiolite Nanocellular Materials with a Bimodal Cellular Structure. Macromol. Mater. Eng. (doi:10.1002/mame.201900041)

Wang G, Zhao J, Mark LH, Wang G, Yu K, Wang C, Park CB, Zhao G. 2017 Ultra-tough and super thermal-insulation nanocellular PMMA/TPU. Chem. Eng. J. 325, 632–646. (doi:10.1016/j.cej.2017.05.116)

Costeux S, Zhu L. 2013 Low density thermoplastic nanofoams nucleated by nanoparticles. Polymer (Guildf). 54, 2785–2795. (doi:10.1016/j.polymer.2013.03.052)

Costeux S, Khan I, Bunker SP, Jeon HK. 2014 Experimental study and modeling of nanofoams formation from single phase acrylic copolymers. J. Cell. Plast. 51, 197–221. (doi:10.1177/0021955X14531972)

Reglero Ruiz JA, Dumon M, Pinto J, Rodriguez-Pérez MA. 2011 Low-density nanocellular foams produced by high-pressure carbon dioxide. Macromol. Mater. Eng. 296, 752–759. (doi: https://doi.org/10.1002/mame.201000346)

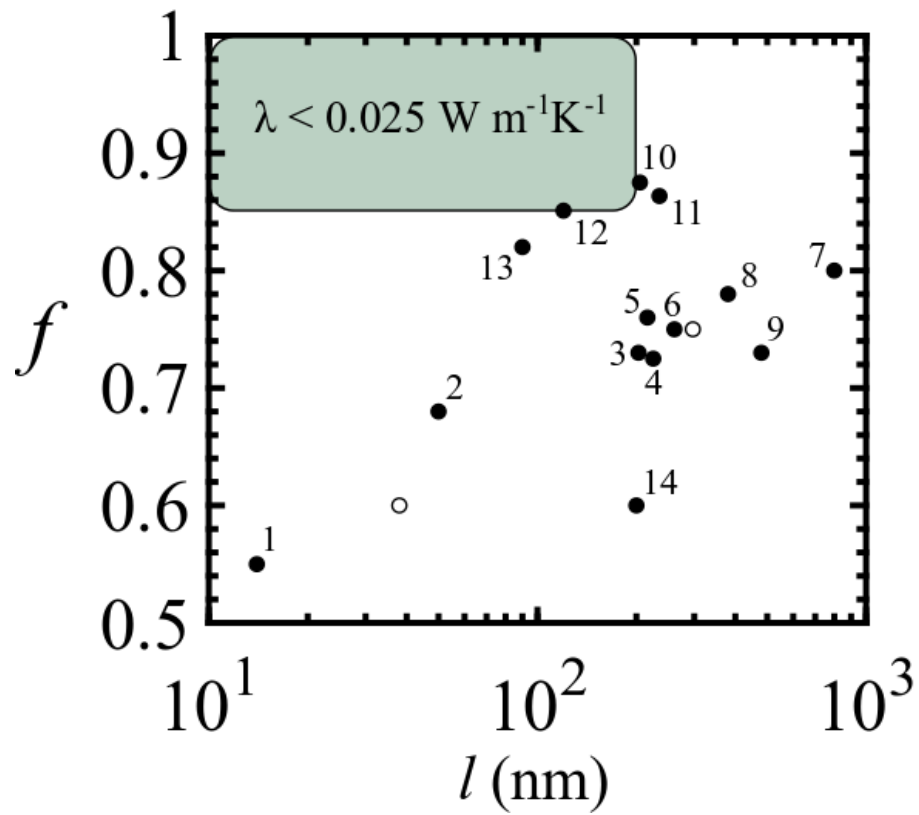

Figure 1 in publication version: Reported porosity $f$ versus void size $l$ data of high porosity (P)MMA-based nanofoams produced via solid-state foaming. The '○' markers refer to results obtained in this work. The '●' markers refer to data retrieved from references in the list above.